\documentclass[aps,floats,showpacs,showkeys,preprintnumbers,nofootinbib]{revtex4}

\usepackage{color}
\usepackage{tipa}
\usepackage{yfonts}
\usepackage{mathrsfs}
\usepackage{eufrak}
\usepackage{centernot}
\usepackage[T1]{fontenc}
\usepackage{egothic}

\usepackage{bbm}                                                  

\newcommand{\nc}{\newcommand}
\nc{\beq}{\begin{equation}}  \nc{\eeq}{\end{equation}}
\nc{\bea}{\begin{eqnarray}}  \nc{\eea}{\end{eqnarray}}
\nc{\baa}{\begin{array}}     \nc{\eaa}{\end{array}}
\nc{\bit}{\begin{itemize}}   \nc{\eit}{\end{itemize}}
\nc{\ben}{\begin{enumerate}} \nc{\een}{\end{enumerate}}
\nc{\bce}{\begin{center}}    \nc{\ece}{\end{center}}
\nc{\bpm}{\begin{pmatrix}}   \nc{\epm}{\end{pmatrix}}
\nc{\bvt}{\begin{verbatim}}  \nc{\evt}{\end{verbatim}}
\def\half{\frac12}	

\def\to{\rightarrow}

\def\boldoverdot{\,{\raise6pt\hbox{\bf.}\!\!\!\!\>}}

\def\then{{\quad\Rightarrow\quad}}

\def\gcal{{\cal G}}

\def\ucal{{\cal U}}

\def\ee{{\bf e}}
\def\ff{{\bf f}}

\def\lll{{\bf l}}

\def\nn{{\bf n}}

\def\ttt{{\bf t}}

\def\vv{{\bf v}}

\def\yy{{\bf y}}

\def\AA{{\bf A}}

\def\CC{{\bf C}}

\def\QQ{{\bf Q}}

\def\mBB{{\mathbbm M}}

\def\oBB{{\mathbbm O}}

\def\rBB{{\mathbbm R}}

\def\uBB{{\mathbbm U}}
\def\vBB{{\mathbbm V}}

\def\zBB{{\mathbbm Z}}
\def\mati{{\mathbbm1}}

\usepackage{bm}
\def\alpbf{{\bm\alpha}}		
\def\betbf{{\bm\beta}}		
\def\gambf{{\bm\gamma}}		
\def\zetbf{{\bm\zeta}}		
\def\kapbf{{\bm\kappa}}		
\def\mubf{{\bm\mu}}		
\def\ssb{spontaneous symmetry breaking}
\def\vev{vacuum expectation value}
\def\irrep{irreducible representation}

\def\tr{ \hbox{tr}}

\def\diag{\hbox{\diag}}
\def\sm{Standard Model}
\def\vevof#1{\left\langle #1 \right\rangle}
\def\ket#1{\left| #1 \right\rangle}

\def\doubleundertext#1{
{\undertext{\vphantom{y}#1}}\par\nobreak\vskip-\the\baselineskip\vskip4pt%
\undertext{\hbox to 2in{}}}
\def\inbox#1{\vbox{\hrule\hbox{\vrule\kern5pt
     \vbox{\kern5pt#1\kern5pt}\kern5pt\vrule}\hrule}}
\def\sqr#1#2{{\vcenter{\hrule height.#2pt
      \hbox{\vrule width.#2pt height#1pt \kern#1pt
         \vrule width.#2pt}
      \hrule height.#2pt}}}

\def\today{\ifcase\month\or
  January\or February\or March\or April\or May\or June\or
  July\or August\or September\or October\or November\or December\fi
  \space\number\day, \number\year}
\def\pmb#1{\setbox0=\hbox{#1}%
  \kern-.025em\copy0\kern-\wd0
  \kern.05em\copy0\kern-\wd0
  \kern-.025em\raise.0433em\box0 }

\def\pmbb#1{\setbox0=\hbox{#1}%
  \kern-.02em\copy0\kern-\wd0
  \kern.04em\copy0\kern-\wd0
  \kern-.02em\raise.03464em\box0 }
\def\up#1{^{\left( #1 \right) }}
\def\inv#1{\frac1{#1}}
\def\su#1{{SU(#1)}}
\def\ui{U(1)}
\def\sumprime_#1{\setbox0=\hbox{$\scriptstyle{#1}$}
  \setbox2=\hbox{$\displaystyle{\sum}$}
  \setbox4=\hbox{${}'\mathsurround=0pt$}
  \dimen0=.5\wd0 \advance\dimen0 by-.5\wd2
  \ifdim\dimen0>0pt
  \ifdim\dimen0>\wd4 \kern\wd4 \else\kern\dimen0\fi\fi
\mathop{{\sum}'}_{\kern-\wd4 #1}}
\font\sanser=cmssq8

\definecolor{straw}{rgb}{1,1,0.50}
\definecolor{lstraw}{rgb}{1,1,0.70}
\definecolor{red}{rgb}{1.00,0.00,0.00}
\definecolor{teal}{rgb}{0.70,0.70,0.90}
\definecolor{darkblue}{rgb}{0,0,0.50}
\definecolor{lblue}{rgb}{0.8,0.8,1}

\def\detail#1{\bigskip
\hspace{-.5in}{\small
\fcolorbox{black}{white}{
\begin{minipage}[h]{6 in}
\noindent\hspace{-.3in}\fcolorbox{black}{teal}
       {\textcolor{darkblue}{\sanser Detail}}
\medskip \par{\tiny #1}
\end{minipage}}
\bigskip
}}

\def\detail#1{}

\def\ket#1{ \left|#1\right\rangle}

\def\dh{{\hat d}}

\def\th{{\hat t}}

\def\tw{t_{\rm w}}

\def\pnb{{\phi_{n,\betbf}}}

\def\bz{{\betbf > 0}}

\def\sw{s_{\rm w}}
\def\At{\tilde A}
\def\AAt{\tilde\AA}

\def\isa{{\bar k}}
\def\sa{\alpbf^\isa}
\def\sw{s_{\rm w}}

\def\tsing{T\up{\mathfrak{s}}}
\def\rot{\mathfrak{r}}
\def\Rot{\mathfrak{R}}
\def\hyp{{\hbox{\egothfamily y}}}

\begin{document}

\preprint{DCP-10-03\cr UCRHEP-T497}

\title{Constraints on realistic Gauge-Higgs unified models}

\author{Alfredo Aranda}
\email{fefo@ucol.mx}
\affiliation{Facultad de Ciencias - CUICBAS, Universidad de Colima, M\'exico
\\ Dual C-P Institute of High Energy Physics, M\'exico}

\author{Jos\'e Wudka}
\email{jose.wudka@ucr.edu}
\affiliation{Department of Physics, University of California,
Riverside CA 92521-0413, USA}

\begin{abstract}
We investigate the general group structure of gauge-Higgs unified models. We find 
that a given embedding of the \sm\ gauge group will imply the presence of 
additional light vectors, except for a small set of special cases, which we 
determine; the arguments presented are independent of the compactification 
scheme. For this set of models we then find those that can both accommodate 
quarks and have a vanishing oblique T-parameter at tree-level. We show that none 
of the resulting models can have $|\sw| \sim1/2 $ (the sine of the weak-mixing 
angle) at tree-level and briefly discuss possible solutions to this problem.
\end{abstract}

\pacs{11.10.Kk, 11.15.-q, 12.10.-g}
\keywords{gauge theories, Higgs boson, extra dimensions, gauge-Higgs unification}

\maketitle

\section{ Introduction}

If there are compact extra dimensions it is then possible to associate 
4-dimensional scalars with the {\it extra} components of gauge bosons; 
these scalars can then be
responsible for the breaking of electroweak (EW) 
symmetry~\cite{Hosotani:1988bm}, playing
the same role as the \sm\ Higgs doublet. In this so-called  
Gauge-Higgs (GH) unification
one first proposes a non-Abelian gauge symmetry in the full 
D-dimensional space time (where ${\rm{D}}=4+n$),
and assumes $n$ of these are compactified, usually over an orbifold
\cite{Hebecker:2001jb} 
(for a review see {\it e.g.} \cite{Quiros:2003gg}).
The compact space together with boundary conditions are then chosen
to insure the presence of light 4-dimensional vector modes 
corresponding to a low-energy
$\su3 \times \su2 \times \ui$ gauge group, together with a 4-dimensional
scalar sector that leads to the right symmetry breaking pattern
to $\su3 \times \ui_{EM}$. Models constructed within this 
scenario involve two high energy scales: 
the compactification radius $L$ and
an ultraviolet (UV) cutoff $\Lambda$,
beyond which the model ceases to be perturbative. This UV cutoff appears
because the models are non-renormalizable;
it is not hard to insure $ \Lambda L \gg 1 $ as required by 
consistency~\cite{Chacko:1999hg,Grzadkowski:2007xm}.

Among the virtues of such a scheme, in addition to the absence of fundamental
scalars, is the possible solution to the hierarchy
problem~\cite{Hatanaka:1998wr,Sakamoto:2006wf}. Another feature is the fact that the low-energy
effective scalar potential, which is responsible for the EW symmetry
breaking, is determined by the group structure, fermion 
content, and boundary conditions of the theory.

Several realizations~\cite{Hall:2001zb,Burdman:2002se,
Haba:2002vc,Gogoladze:2003bb,Scrucca:2003ut,Haba:2004qf,
Haba:2004jd,Chiang:2010hy,lim:2007xh} of this mechanism have been achieved with various
 degrees of success. 
5-dimensional models generally predict a very light Higgs boson. This is due to
the absence of a tree-level quartic term in the scalar 
potential~\cite{Haba:2004bh,Gogoladze:2007ey}, which can be ameliorated by 
considering models with 6 (or more) dimensions. 
These models also typically predict a tree-level value of
$ \sin^2 \theta_{\rm w} $ ($\theta_w$ denotes the weak mixing angle)
higher than the experimental observation of $ \sim1/4 $;
in fact, for 5-dimensional gauge theories, it has been 
shown~\cite{Grzadkowski:2006tp} that any model that assumes a low-energy \sm (SM) 
gauge boson spectrum only, is inconsistent with the requirements $\sin^2\theta_W 
\simeq 1/4$ and $\rho \equiv m_W^2/(m_Z^2\cos^2\theta_W)=1$
at tree level (equivalently, a vanishing tree-level
oblique T-parameter). This problem has been addressed
by introducing additional  $\ui$ 
factors~\cite{Antoniadis:2001cv,Scrucca:2003ra,Cacciapaglia:2005da} and/or brane 
kinetic terms~\cite{Burdman:2002se,Agashe:2004rs,Aranda:2005ze}. 

In this paper, following the analysis presented in~\cite{Grzadkowski:2006tp}, we 
determine to what extent GH models naturally satisfy the following
set of low-energy constraints:
\bit
\item The only gauge bosons with masses at the EW mass scale are those
present in the \sm.
\item $\rho = 1$ (at tree-level), {\it i.e.} we look for models that contain only 
isodoublet 
or isosinglet 4-dimensional scalars~\footnote{While isodoublets are not the only 
``$\rho$-safe'' representations, they {\em are} the only ones with this property 
that occur in the models being considered.},
\item Contain representations that can accommodate all \sm\ particles;
in particular at least one representation should contain 
isodoublet states of hypercharge $1/6 $, corresponding to the 
left-handed quark fields.
\item Lead to acceptable values of $\sw^2$.
\eit
For all simple groups we determine whether or not these 
requirements can be fulfilled by the group structure itself without allowing 
additions to the models such as fundamental scalars or brane couplings. We 
present the analysis and results for an arbitrary number of extra dimensions and 
all compactification schemes. Our approach is group theoretical supplemented by 
the above four experimentally-motivated conditions.

We find that it is relatively easy to satisfy the first two requirements, 
while the third one is obeyed in only a relatively small set of models
(listed in eq. \ref{eq:models}). It proved impossible, in addition, to
satisfy $ \sw^2\sim1/4 $:
in the absence of brane couplings all GH models must either contain
additional light gauge bosons (with masses $\sim m_{\rm W,Z}$), 
or must have a tree-level weak-mixing angle 
substantially different form the observed one, or cannot accommodate quarks.
The masses of the additional
light gauge bosons are of the same order as those of the $W$ and $Z$ and
are subject to similar radiative corrections, so we do not expect
them to be split form the electroweak scale by loop effects;
hence models that exhibit
such particles appear phenomenologically excluded. One of the main results of 
this paper is to provide a simple way of determining whether these 
undesirable states are present without a reference to the compactification
scheme.

In contrast, radiative corrections  to weak-mixing angle can be substantial and 
can easily lead to a 50\%\ (or larger) 
effect~\cite{Dienes:1998vh,Dienes:1998vg}. For this 
reason we will first determine gauge groups that comply with the first three 
conditions above and then discuss the possibility of obtaining a 
phenomenologically viable weak-mixing angle at the electroweak scale.

In this paper we concentrate on the gauge-boson structure of the theory; fermions 
will be discussed only in connection with their possible $\su2\times\ui $ quantum 
numbers. The possibility of creating specific models that accommodate the right 
light fermion with the correct $\su3\times\su2\times\ui$ quantum numbers, and - 
more challenging - the observed Yukawa couplings, will not be studied here. It is 
known that, even if absent at tree-level, radiative contributions generate 
4-dimensional couplings on the orbifold fixed points~\cite{Georgi:2000ks}. Given a 
specific realization of the GH scenario one must take this into account, and 
determine the extent of the associated effects on the low-energy theory -- 
especially if it is assumed that the \sm\ fermions are localized in one of these 
fixed points . This investigation requires a specific choice of model (including 
gauge field, fermion representations, orbifold compactification and periodicity 
conditions on the fields) and lies beyond the scope of the model-independent 
restrictions presented here. 

The paper is organized as follows: after a brief description of the conventions 
used in our analysis, we present the general gauge and space time setup in Section~\ref{s2}. 
The gauge transformations and Kaluza-Klein decomposition of the fields, as well as 
the analysis regarding light vector bosons and 
the necessary conditions for their absence are presented in Section~\ref{s3}. 
In section~\ref{s5} we present the 
vector boson mass matrix and the requirements needed for matter fields. Finally 
we present our results in section~\ref{s6}. We have also included a couple of 
appendixes at the end with details of the calculations. 

\section{ Lagrangians and symmetries} \label{s2}

The material presented in this section is not new, it is included
for convenience and in order to introduce the notation that will 
be used throughout the paper.

\subsection{Conventions}
In describing the group structure of the models being considered we
find it convenient to use a canonical basis where we
denote the Cartan generators by $ C_i $
(or, when appropriate as a vector \CC), and the root generators by
$ E_\betbf $. These satisfy
\beq \label{eq:generatoralgebra}
[ \CC, E_\betbf] = \betbf E_\betbf\,; \quad
[E_\betbf,E_{\-\betbf} ] = \betbf \cdot \CC \,;\quad
[E_\betbf, E_\gambf ] = N_{\betbf, \gambf} E_{\betbf + \gambf} ~~
(\betbf + \gambf \not= {\bf0})
\eeq
where we will not need the explicit form of $N_{\betbf, \gambf}$
\cite{Gilmore:1974}.
Note that in this basis the structure constants are not
completely antisymmetric.

We will denote the simple roots by
$ \alpbf^i $ and the corresponding fundamental weights
 by $ \mubf_j $ (see \cite{Gilmore:1974,Georgi:2000ks} for details), with
\beq
\alpbf^i \cdot \mubf_k = \half \left| \alpbf^i \right|^2 \delta_{ik}
\then \mubf_k = \sum_j \left(a^{-1}\right)_{j k} \alpbf^j
\eeq
(no sum over $i$ in the first expression) where
\beq
 a^{ij} = 2 \frac{\alpbf^i\cdot\alpbf^j }{\alpbf^i\cdot\alpbf^i }
\label{eq:cart.mat}
\eeq
(no sum over $i$) 
is called the Cartan matrix (which is not symmetric in general).
We also find convenient
to introduce the rescaled weights
\beq
\tilde\mubf_j = \frac2{|\alpbf^j|^2} \mubf_j
\then
\alpbf^i \cdot \tilde\mubf_j = \delta_{ij}
\label{eq:def.of.mu}
\eeq
(no sum over $j$ in the first expression) .

For the adjoint representation we assume that the
corresponding matrices are normalized according to
(using the same symbol for the generator as for its
representation)
\beq
{\rm tr} C_i C_j = \delta_{i,j} \,, \quad
{\rm tr} E_\alpbf E_\betbf = \delta_{\alpbf+\betbf, 0} \,, \quad
{\rm tr} E_\alpbf C_i =0 \,,
\label{eq:norm}
\eeq
which ensures that $ (-1/4) {\rm tr}
|\partial_\mu A_\nu - \partial_\nu A_\mu|^2,~
A_\mu = \sum_i A_\mu^i C_i + \sum_\betbf A_\mu^\betbf E_\betbf $,
is properly normalized.

When no confusion will arise,  we will
refer to a gauge field and it associated generator interchangeably;
for example we will often refer to a
generator as ``corresponding'' to a light vector boson, by which
we imply that the gauge field associated with that generator
will have a light mode.

\subsection{Compactifications and the space group $ \Gamma$}
We consider theories defined on a $4+n$ dimensional
space time of the form $ \mBB \times (\rBB^n/\Gamma) $ where
$\mBB$ denotes the usual 4-dimensional Minkowski space
and $ \Gamma $ denotes a discrete group~\cite{Hosotani:1988bm} 
(for a review see \cite{Quiros:2003gg}) whose elements
$ \gamma \in \Gamma $ act on $ \rBB^n$ as follows~\footnote{The
notation is borrowed from solid state literature, for an accessible
introduction see \cite{Koster:1956}.}:
\beq
\gamma = \{ \rot|\lll\} \then \{ \rot|\lll\} \yy = \rot \yy + \lll \ ,
\eeq
where the $\rot$ are $ n \times n $ orthogonal matrices,
\yy\ denote the coordinates of $ \rBB^n$, and the $ \lll$
are $n$-dimensional translation vectors
(the notation is the same as
the one used in solid state for crystal groups).
We assume that $ \Gamma $ acts trivially on $ \mBB $:
$ \gamma x^\mu = x^\mu $ where $ x^\mu$ denote the coordinates of $ \mBB $.

The multiplication rule for the elements in $ \Gamma $
can be easily derived from their action on \yy,
for example
\beq
\{ \rot'|\lll'\} \{ \rot|\lll\} = \{\rot'\rot | \rot'\lll + \lll'\} \,;
\quad \{ \rot|\lll\}^{-1} =  \{ \rot^{-1} | - \rot^{-1} \lll\} \ .
\eeq

The group $ \Gamma $ is assumed to have an
(Abelian) translation subgroup
$ \Theta $  composed of all elements of the form
$ \{ \mati | \ttt \} $, where the translations vectors \ttt\
are linear combinations of a
set of basis vectors $ \{ \ttt_ i \} $ with integer coefficients,
\beq
\Theta = \left\{ \{ \mati |\sum k_i \ttt_i\}, ~k_i={\rm integer}
 \right\} \,;
\eeq
note that, in general, the vector
\lll\ in $ \{\rot|\lll\}$ need not be a translation when
$ \rot\not=\mati$. Using the multiplication
rule we find that
\beq 
\{ \rot|\lll\}^{-1} \{ \mati |\ttt\} \{ \rot|\lll\}= \{ \mati |\rot^{-1} \ttt\}
\then \rot^{-1} \ttt \in \Theta \ .
\eeq
It follows that $ \Theta $ is an invariant subgroup, and that
for all rotations $\rot$ and translations \ttt\ the
vector $ \rot \ttt$ is also a translation.

\subsection{Gauge-field Lagrangian and automorphisms}
We denote gauge vector fields by $A^M_a = (A^\mu_a,A^m_a)$ with the Greek
indices associated with non-compact directions and Latin lower case
indices with compact directions; we often write $ \AA_a =
(A^{m=1}_a, \ldots, A^n_a)$.

We assume the following action of $ \Gamma $ on the $A^M_a$:
\bea
A^\mu_a(x,\yy') &=& \vBB(\gamma)_{a b} A^\mu_b(x,\yy) \,; \qquad
\yy' = \gamma\yy =  \{ \rot|\lll\} \yy = \rot \yy + \lll \cr
\AA_a(x,\yy') &=&  \vBB(\gamma)_{ab} \Rot(\gamma) \AA_b(x,\yy)
\label{eq:qaction}
\eea
for all $ \gamma \in \Gamma $, where the matrices $\Rot$ act on the indices
associated with the compact directions:
$ (\Rot\AA_b)^m = \Rot^m{}_l A_b^l $.

Defining the curvature tensor by~\footnote{The structure constants obey
$
f_{a b c} = - f_{b a c}
$
but, in general $ f_{a b c} + f_{a c b } \not =0 $
(though there are bases where this does hold)}
\beq
F^{MN}_a = \partial^M A^N_a - \partial^N A^M_a + g f_{b c a} A^M_b A^N_c \ ,
\eeq
we find that the $F^2$ term in the Lagrangian will
be invariant under (\ref{eq:qaction}) provided
\bea
\rot = \Rot(\gamma); ~ \gamma= \{ \rot | \lll \} && \Rot(\gamma){}^T \Rot(\gamma) = \mati \ , \cr
f_{a'b'c'} = f_{abc} \vBB(\gamma)_{a a '} \vBB(\gamma)_{b b'}
\vBB^\dagger(\gamma)_{c' c} &&
\vBB(\gamma){}^\dagger \vBB(\gamma) = \mati \ ,
\label{eq:autom}
\eea
so that $ \vBB$ must be an automorphism of the
gauge group $ \gcal $~\cite{Hebecker:2001jb}.
Note also that
the equality $ \Rot(\gamma) = \rot $ implies $\Rot(\gamma)$ is
independent of the vector $ \lll $. In  particular
\beq
\Rot(\{ \mati | \ttt \} ) = \mati \ .
\label{eq:R.under.T}
\eeq

\section{KK expansions} \label{s3}

\subsection{Consequences of covariance under translations}
The matrices $ \vBB(\{\mati|\ttt\})$ carry a representation of $\Theta
$. Since $ \Theta $ is an Abelian group it has only one-dimensional
\irrep s, so we can choose a basis where $ \vBB(\{\mati|\ttt\})_{a b} =
v_a(\ttt) \delta_{a b} $ for all \ttt. Since the vectors \ttt\ are of the
form $ \sum_i k_i \ttt_i$ for some integers $k_i $,
and since $ \{ \mati | \ttt_i \} \in \Gamma$, then
\beq
v_a(\ttt) = \prod_i
\left[ v_a(\ttt_i) \right]^{k_i} \ ,
\eeq
while the unitarity of the representation requires
\beq
v_a(\ttt_i)= e^{ i c_{a i}}, \quad c_{a i} \rightarrow \hbox{real},
~~ |c_{a i} | < \pi \ .
\eeq
In this basis, using (\ref{eq:qaction}) and (\ref{eq:R.under.T}),
\beq
A_a^N(x, \yy+ \ttt) = e^{i \sum_i k_i c_{a i}} A_a^N(x,\yy) \ .
\eeq

Let $ \{ \kapbf_j\} $ be a set of linearly independent vectors
dual to the $ \ttt_i$, that is, $ \kapbf_j \cdot \ttt_i = \delta_{ij} $.
Then we can expand
\bea
A^N_a(x,\yy) &=& e^{ i \QQ_a \cdot \yy}
\sum_\nn e^{2 \pi i \yy \cdot \sum_j n_j \kapbf_j} \At^N_a(x,\nn); \quad
\QQ_a = \sum_j c_{a j} \kapbf_j \ ,
\label{eq:KK.expansion}
\eea
which corresponds to the usual expansion in Kaluza-Klein
(KK) modes~\cite{Hosotani:1988bm,Quiros:2003gg}. 
It follows that $ \At^N_a(x,\nn) $ will have a
the tree-level mass $ \sim | \QQ_a + 2 \pi \sum n_j
\kapbf_j | $;  we will assume that the scales associated with the $
\kapbf_i $ are large compared to the electroweak
scale. In this case
all light modes $ \At^N_a $ must correspond to $ \nn =0 $
and $ \QQ_a =0 $. But $ \QQ_a =0 $ requires $ c_{a i } = 0 $,
which implies that gauge fields that have light modes are
translationally invariant.

Since $ \Theta $ is an invariant subgroup we find that
for all $ \gamma$,
$  \vBB(\gamma )_{ab} =0 $ {\em unless } $\QQ_a = \QQ_b $.
In particular the $ \vBB $ do not mix the light 
and heavy KK modes. Using this and substituting
(\ref{eq:KK.expansion}) in (\ref{eq:qaction})
we find that the light modes must obey
\beq
\vBB(\gamma)_{a b} \At^\mu_b(x,0)
=
\At^\mu_a(x,0) \qquad
\vBB(\gamma)_{a b} \Rot(\gamma) \AAt_b(x,0)
=
\AAt_a(x,0) \ ,
\label{eq:mzero}
\eeq
that is, light 4-dimensional vectors are associated with the ``trivial''
subspace where $ \vBB= \mati $, while light 4-dimensional scalars are
associated with the subspace where $ \vBB \otimes \Rot = \mati $.

One of the challenges in constructing a realistic theory
of this type is to find a ``compactifying'' group $ \Gamma $,
a gauge group $ \gcal $ and representations $ \vBB$
and $ \Rot $ such that the solutions to (\ref{eq:mzero})
will correspond to the bosonic sector of the \sm\ (possibly
with an extended scalar sector). While it is always
easy to  ensure that a given field has a light mode,
the relations (\ref{eq:autom}) often imply the presence
of  {\em additional} light vectors which, as we will see,
have masses of the same order as the \sm\ gauge bosons.
Requiring the absence of such undesirable light particles
severely restricts the choice of gauge groups $ \gcal $.

The gauge transformations are
\beq
A^N \to U^\dagger \left( i \partial^N + A^N \right) U; \qquad
A^N = g A^N_a T_a\,, ~~ U \in \gcal
\eeq
equivalently,
\beq
A^N_a \to \ucal^N_a + \uBB_{ab} A^N_b;
\qquad
T_a \ucal^N_a  = i U^\dagger \partial^N U,
\quad
T_a \uBB_{ab} = U^\dagger T_b U \ .
\label{eq:gauge.transf}
\eeq

Consider now those $ \ucal,~\uBB$ that depend only on $x$. Using
(\ref{eq:KK.expansion}) in (\ref{eq:gauge.transf}) implies
that $ \ucal^\mu_a \not=0 $ if $ \QQ_a =0 $; and
$ \uBB_{ab}(x) \not=0 $ requires $
\QQ_a - \QQ_b = 2 \pi \bar n_j\up{a b} \kapbf_j$
with $ \bar n_i\up{a b}  $ integers;
in particular if $ \QQ_a =0 $, then either $ \QQ_b =0 $
or else $ \uBB_{ab}(x) =0 $.
Then, for the light fields obeying (\ref{eq:mzero}),
\bea
\At^\mu_a(x,{\bf0}) &\to& \ucal^\mu_a(x)+ \sum_b
\uBB_{ab}(x) \At^\mu_b(x,{\bf0}) \cr
\AAt_a(x,{\bf0})
&\to& \sum_b \uBB_{ab}(x)
 \AAt_b(x,{\bf0})
\eea
showing that when $ \ucal^\mu_a \not =0 $ the
$ \At^\mu_a(x,\nn=0) $ transform as gauge fields
and, because of (\ref{eq:mzero}) these  also
transform trivially under $ \Gamma $.
We denote these as the light gauge fields, and the set of
\yy-independent transformations as the light gauge group $G$.
The light gauge group $G$ is a subgroup of $\gcal$, so that all
fields can be classified according to their $G$ representation.
For a realistic theory~\footnote{Models with additional $\ui$
factors appear frequently, but in the absence of brane couplings
or bulk-propagating scalar fields,
the masses of the additional vector bosons are unacceptably low.}
 $ G = \su3 \times\su2 \times \ui $.

As in the 5-dimensional case~\cite{Grzadkowski:2006tp} the $\su2$ generators
$ J^{0,\pm} $ are specified
by choosing a root $ \alpbf$, while the hypercharge
generator $Y$ can be taken
as a linear combination of Cartan generators:
\beq
J^0 = \inv{|\alpbf|} \hat\alpbf\cdot\CC \,, \quad
J^\pm =  \frac{\sqrt{2}}{|\alpbf|} E_{\pm\alpbf} \,;
\qquad Y = \hyp \cdot \CC \,.
\eeq
Though $ \alpbf $ can be any root, there is always a freedom to relabel
axes and axis directions, so that many choices are
equivalent. A straightforward inspection of 
the various classical groups 
shows that, in fact, $ \alpbf $ can be taken to be one of
the simple roots.

Using these definitions we can find the isospin $s$ and
$z$-component isospin $s_z$ of any root $ \betbf$:
\beq
s = \frac{n_+ + n_- }2 \,, \qquad
s_z = \frac{n_- - n_+ }2 \ ,
\label{eq:def.of.s.sz}
\eeq
where $n_\pm $ are non-negative integers such that
$\betbf + k  \alpbf$ and $-n_n \le k \le n_+ $ are roots but
$\betbf \pm (n_\pm +1) \alpbf $ are not.
If $ \betbf $ is also a simple root then $n_- =0 $
so that $ s = - s_z = n_+/2 $.

The vector \hyp\ is unspecified, except for the requirements that
the model should contain isodoublets of hypercharge $1/2 $ that can acquire
a \vev, and that $ J^{0,\pm}$ commute with $Y$:
\beq
\left[ J^\pm , Y \right]  =0 \then \hyp\cdot\alpbf =0
\eeq

Concerning the light 4-dimensional scalars we require that they
give rise to  $ \rho = 1$ at tree level. Since these scalars are
associated with the adjoint generators their isospin can be read-off
form  the Cartan matrix of the gauge group $ \gcal $; for the
simple groups a direct examination shows that only isospin $ \le 3/2 $
will occur (see appendix~\ref{sec:class.groups}), and so this constraint by 
itself does not rule out any of them. 
The scalars are then taken to be associated with specific root
generators $ E_\betbf $ and a
realistic model must ensure that only the isospin $1/2 $
modes can acquire a \vev.



Requiring consistency between (\ref{eq:qaction}) and (\ref{eq:gauge.transf})
implies
\bea
&& \ucal^\mu_a(x,\yy') = \vBB_{ab}(\gamma) \ucal^\mu_b(x,\yy) \qquad
\ucal^m_a(x,\yy') = \Rot^m{}_l \vBB_{ab}(\gamma) \ucal^l_b(x,\yy) \cr
&& \uBB(x,\yy') = \vBB(\gamma) \uBB(x,\yy) \vBB^{-1}(\gamma) \ ,
\eea
where $ \yy' = \rot \yy + \lll,~~ \gamma=\{\rot|\lll\} $. In
particular for $ \uBB \in G$, $ \uBB $ is independent of $x$
and
\beq
\left[ \uBB(x) , \vBB(\gamma)\right] = 0 \ .
\eeq
This relation implies (by Schur's lemma) that such $ \uBB$ do
not mix $ \Gamma $ \irrep s carried by the $ \vBB $, and the
$ \vBB $ do not mix $G$ \irrep s carried by the $ \uBB $. In
particular, the $ \vBB $ will not mix generators that
have different $G$ quantum numbers.

Consider now the set of generators that are $G$ singlets, denoted
by $ \tsing_S$, and a set of generators $\tilde T_r $ that have
fixed $G$ quantum numbers (in our case, fixed hypercharge,
isospin and $z$-isospin component). Then it follows that
$ \vBB_{S a} =0 $ unless $ a = R $, corresponding to some $ \tsing_R$,
and $ \vBB_{r a} = 0 $ unless $ a = s $, corresponding to some
$ \tilde T_s $. The automorphism condition then implies
(using a  basis where
$ f_{abc}$ is antisymmetric in $a$ and $b$ but not in all 3 indices)
\beq
 f_{T r s} = \vBB_{T' T} \vBB_{r' r}
\vBB^{-1}_{s s'}  f_{T' r' s'}
\then \sum_r  f_{T r r} = \sum_{T'} \vBB_{T' T} \sum_{r'}
 f_{T' r' r'} \ .
\eeq
Then
\beq
 F_S =  F_R \vBB_{R S} \,, \quad
\hbox{\rm where} \quad
 F_S = \sum_r  f_{S r r} \ .
\eeq

Now define
\beq
\tilde \tsing = \sum_S  F_S \tsing_S
 = \sum_{S\,r}  f_{S r r} \tsing_S \ ,
\eeq
which is an $G$ singlet, while under $ \Gamma $
\beq
\tilde\tsing \to \sum \vBB_{ R S} \tsing_S  F_R =
\tilde\tsing \ ,
\eeq
so that this generator is also a $ \Gamma $ singlet.
This generator depends on the choice of $ \tilde T_r $ so that
there will be a $ \tilde\tsing$ for each set of $G$ quantum numbers.
Note that any linear combination of the $ \tilde\tsing $
will also be a $G$ and $ \Gamma $ singlet.

Now, the expression for $ \tilde\tsing $ involves $  f_{S r r} $,
which is not zero only if the commutator $ [\tsing_S , \tilde T_r ] $
has a term proportional to $ \tilde T_r $ itself. Looking now
at the commutators in terms of roots and Cartan generators it is
clear that this can happen only if $ \tsing_S $ is a linear combination
of Cartan generators and $ \tilde T_r $ is a root generator:
\beq
\left[ \vv \cdot \CC, E_\betbf \right]  = \vv \cdot \betbf E_\betbf
\then f_{\vv\cdot\CC \, E_\betbf \, E_\betbf} = - i \vv \cdot \betbf \ ,
\eeq
so we identify $ \tsing_S $ with $ \vv \cdot \CC $, which
will be an $ \su2$ singlet provided
\beq
\vv\cdot\alpbf =0 \ .
\eeq
It is also a $\ui$ singlet since the
hypercharge generator is of the form $ \hyp \cdot \CC $.

Consider now a series of vectors $ \hat \vv_S $ perpendicular to
$ \alpbf $ and satisfying
\beq
\sum_S \hat\vv_S  \otimes \hat\vv_S = \mati - \hat\alpbf \otimes \hat\alpbf \ .
\eeq
Then we can take $ \tsing_S = \hat\vv_S \cdot\CC$ and
the $G$ and $ \Gamma $ singlet generators are of the form
\beq
\sumprime_\betbf
\sum_S\left(\hat\vv_S \cdot \betbf\right)  \hat\vv_S \cdot \CC
= \left(\sumprime_\betbf \betbf_\perp \right)\cdot \CC \ ,
\label{eq:sum.1}
\eeq
where $ \betbf_\perp = \betbf - (\hat\alpbf \cdot \betbf) \hat\alpbf$, and
the prime indicates that the sum is over all roots with
a specific set of $G$ quantum numbers. In the following we write
\beq
\sumprime_\betbf \betbf_\perp = \hyp'_q,~ ~
q = \{ h ,s,s_z\}
\label{eq:sum.2}
\eeq
Since any two roots $ \betbf,~\betbf' $
in the same $ \su2$ multiplet satisfy  $ \betbf = \betbf'
+ n \alpbf  $, it follows that $ \betbf_\perp = \betbf'_\perp $.
If $ \betbf,~ \betbf'$ carry $G$ quantum numbers
$q = \{h, s, s_z \},~
q' = \{h, s, s_z' \} $ respectively, then 
$ \hyp'_q = \hyp'_{q'} $. Note also that 
\beq
\hyp'_q = - \hyp'_{\bar q} \,, \quad q = \{h , s , s_z \}, ~~
\bar q = \{ -h, s, -s_z \} \,,
\eeq
so that we can restrict ourselves to vectors $ \hyp' $ associated
with positive hypercharge, $ h> 0 $,  roots.

The generators $ \hyp'_q \cdot \CC $
are {\em necessarily} $G$ and $ \Gamma $ singlets~\footnote{
It is possible to follow the same argument
when $ G = \su2 $, the hope being that one
of the $\hyp'_{s,s_z}$ can be used as \hyp. This case,
however, is trivial since all the $ \hyp'_{s,s_z}$
necessarily vanish (see~\ref{sec:ap2}), 
which is related to the fact that
$\su2 $ has no complex representations.}: given a choice of $ \gcal $ and an
embedding of $G$ it is impossible to 
have a light gauge group of rank 4, unless
all the vectors $ \hyp_q' $ are proportional to $ \hyp $.
The fact that a given choice of the \sm\ group as a
subgroup of $ \gcal $ will in general imply the presence
of additional light vector bosons, independently of the
compactification details of the model, is one of the
central results of this paper.

The vectors $\hyp'_q$ need not be linearly independent,
nor do they have to be independent of \hyp. Still,
by taking appropriate linear combinations we can find
and  orthonormal subset
$ \hat\hyp_r, ~ r=0,1, \ldots , R $ with $ \hat\hyp_0 = \hat\hyp $ and
$\hat\hyp_r \cdot \hat \hyp_s = \delta_{rs} $. The generators
$ \hat\hyp_r \cdot \CC $ also are $G$ and $ \Gamma $ singlets,
each generating a $\ui $ subgroup that, by (\ref{eq:mzero}),
corresponds to a light neutral vector boson. This
result is {\em independent} of the choice
of $ \Gamma $ and the representations $ \vBB$ and $ \Rot $;
number $R$ of additional light vector bosons
is determined solely by the gauge structure of the
theory~\footnote{
This refers to the smallest number of light vector bosons; 
specific models, of course, may have additional ones;
for example, if $ \vBB(\gamma ) = \mati $
for all $ \gamma $, {\em all} $A^\mu_a$ will have light
modes.}.

We will show below that these $R$ vector
bosons may acquire masses thorough \ssb, but these are
of the order of the $W$ and $Z$ vector boson masses; in
particular models with $R \ge 1 $ are excluded phenomenologically.
One must therefore choose $G$ (that is,
hypercharge generator $Y = \hyp\cdot\CC$
and $ \alpbf $) and $ \gcal $ such that $R=0 $, which proves to be
a stringent constraint on $ \gcal $ and the embedding of $G$.

\subsection{Necessary conditions for the absence of undesirable singlets}

\label{sec:nec.con}

In this section we list several requirements that gauge-Higgs unified theories
must meet in order to be phenomenologically viable
(at tree level). These
conditions are derived in appendix \ref{sec:group.theo}; here we
only list the results that are relevant to the rest of the paper.

We begin by noting that we can always assume that all the simple
roots have non-negative hypercharge and that
the root $ \alpbf $, which defines
the \sm\ $\su2$ subgroup, is one of the simple roots
(see appendix~\ref{sec:ap1}),
\beq
\alpbf = \sa \ .
\eeq
We can now divide the simple roots into 3 categories
{\it(i)} Those with positive hypercharge, $ \alpbf^{i_k}: ~~
\hyp\cdot \alpbf^{i_k}  = h^k > 0 $. {\it(ii)}
Those, like $ \sa$, with zero hypercharge and non-zero isospin,
$ \gambf^r: ~~
 \hyp \cdot\gambf^r =0,~ \alpbf\cdot\gambf^r \not= 0$
(for $ \gambf^r \not= \sa $).
{\it(iii)} Those that are $G$-singlets: $\zetbf^l: ~~
\hyp \cdot\zetbf^l = \alpbf\cdot\zetbf^l =0$.
It then follows that
\beq
\hyp = \sum h^k \tilde\mubf_{i_k} \ .
\eeq

We have argued above that in order not to have additional 
light vector bosons we must have $ \hyp'_q \propto \hyp $ for all $q$.
A straightforward application of Lie algebra theory (see appendix~\ref{sec:ap3})
shows that necessary conditions for this to occur are:
\bit
\item All simple roots must be either
isodoublets~\footnote{Strictly speaking the condition is
for all non-isosinglets roots to transform according to
the same $\su2$ representation; but we noted earlier that
these representations must have isospin $1/2,~1$ or
$3/2 $, of which only the first produce $ \rho= 1$ at
tree-level.} or isosinglets. This can be satisfied for any simple group.
\item All the isosinglets must have zero hypercharge.
\item All the isodoublets must have the same non-zero hypercharge:
\beq
\hyp = h \sum_{\rm isodoublets} \tilde\mubf_{i_k} \ .
\label{eq:hyp.vec}
\eeq
In particular, there should be no roots of the type $ \gambf^r$
except $ \alpbf $ itself.
\eit
These conditions are useful in that they eliminate a large
number of groups; still, even when met, the low-energy spectrum must
be derived explicitly in order to insure the absence of light $Z'$ 
vectors.
It is worth pointing out that a choice of the $\su2$ subgroup
determines the set of simple roots that transform as isodoublets,
which in turn determines the specific embedding of the \sm\
$\ui$ in $ \gcal $ given by (\ref{eq:hyp.vec}). Hence the $\hyp'$
must be obtained for each choice of $ \alpbf $ in order to determine
the viability of a given model.

\section{The weak mixing angle and hypercharge}\label{s5}

\subsection{Vector-boson mass matrix}

The (canonically normalized) light vector bosons
correspond to the zero modes of the gauge fields associated
with the generators $\hat\alpbf \cdot\CC $, $E_{\pm\alpbf} $
and $ \hat\hyp_r \cdot \CC $, where $ \{ \hat\hyp_0
, \ldots , \hat\hyp_R \} $
denote the orthonormal basis for the subspace generated by $ \hyp$ 
introduced in eq.(~\ref{eq:sum.2} (when there are no additional
$Z'$ bosons $R=0$ and $ \hat\hyp = \hat\hyp_0$);
we denote the corresponding zero modes 
by $ W^0,~W^\pm $ and $B\up r$ respectively. 
Following~\cite{Grzadkowski:2006tp}, but
allowing for the presence of more than one $\su2$-singlet
gauge boson, we expand
\bea
A_\mu  &=& W_\mu^+ E_\alpbf + W_\mu^- E_{-\alpbf} + W_\mu^0 \hat\alpbf\cdot\CC
 + \sum_{r=0}^R B\up r_\mu \hat\hyp_r \cdot\CC + \cdots \cr
A_n &=& \sum_\bz \left( \pnb E_\betbf + \pnb^* E_{- \betbf} \right)
+ \cdots
\eea
where the sum over $ \betbf$ is over those fields such that $
\vevof\pnb \not =0 $, and the ellipsis denote fields with
masses of the order $ | \kapbf | $ as discussed in section
\ref{s3}.

Then, using (\ref{eq:norm}),
\bea
-\tr \left[ A_\mu , A_n \right]^2 &=& \sum_{\bz;
~{\rm isodoublets}} \left| \pnb \right|^2
\left\{ \half \alpbf^2 W^+ \cdot W^- + \left(W_\mu^0 \hat\alpbf \cdot
\betbf + \sum_{r=0}^R B\up r_\mu \hat\hyp_r \cdot \betbf \right)^2 \right\} \cr
&& \quad +  \sum_{\bz;
~G-{\rm singlets}} \left| \pnb \right|^2
 \left(\sum_{r=0}^R B\up r_\mu \hat\hyp_r \cdot \betbf \right)^2 \ .
\label{eq:vbm}
\eea
It is shown in appendix \ref{sec:ap2} that
if $E_\betbf$ is a singlet under $G$, then
$ \hat\hyp_r \cdot\betbf =0 $, so that the $G$-singlet contribution
to the mass term vanishes, and the mass term is simply
\beq
\sum_{\bz;
~{\rm isodoublets}} \left| \pnb \right|^2
\left\{ \half \alpbf^2 W^+ \cdot W^- + \left(W_\mu^0 \hat\alpbf \cdot
\betbf + \sum_{r=0}^R B\up r_\mu \hat\hyp_r \cdot \betbf \right)^2 \right\} \ .
\eeq

This shows that when $R\ge1 $ there will be additional vector bosons
with mass of the same order as that of the $W$ and $Z$.  In particular,
such models contain no mechanism through which the 
tree-level mass of the additional
vectors can be pushed above the experimental limits, {\em irrespective
of the compactification scheme or of the fermion content of the
theory}~\footnote{At least as long as the model is weakly coupled at low energies.}.
This could be corrected by introducing scalars (either in the usual way
or using antisymmetric tensor fields), but in this case the motivation
and the attraction of this type of theories largely disappears.

For the case where $ R=0 $ we assume that the effective potential
for the 4-dimensional scalars insures these get a \vev\ that
preserves the charge generator $ Q = J^0 + Y $, that is
\beq
\vevof\pnb \not=0 \then \hyp\cdot\betbf =\half , \quad
s = \frac{\alpbf\cdot\betbf}{|\alpbf|^2} = - \half \ ,
\label{eq:sz.and.h}
\eeq
which fixes the normalization of \hyp. 
Writing $ B\up0 = B,~\hyp_0 = \hyp $,
and using $ \hat\alpbf \cdot\betbf = - |\alpbf|/2 $,
we find
\beq
\sum_\bz \left| \pnb \right|^2
\left\{ \half \alpbf^2 W^+ \cdot W^- + \inv4 \alpbf^2
\left(W_\mu^0 - \inv{|\alpbf| |\hyp|} B_\mu \right)^2 \right\} \ .
\eeq
From this we read off the tangent of the weak-mixing angle, $\tw$:
\beq
\tan(\theta_{\rm w}) = \tw = \inv{|\alpbf| |\hyp|} \ .
\label{eq:tw}
\eeq
The properly-normalized
$Z$ boson field is then $ Z_\mu = \cos \theta_{\rm w}\, W_\mu^0 
- \sin \theta_{\rm w} \, B_\mu$, and the vector-boson masses are given by
\beq
M_{\rm W}^2  =  \frac{\alpbf^2}2 \sum_\bz \left| \pnb \right|^2
\,, \qquad
M_{\rm Z}^2  =  \frac{ \alpbf^2}{2 \cos^2 \theta_{\rm w}} \sum_\bz \left| \pnb \right|^2
\eeq
so that indeed $ \rho = M_{\rm Z} \cos \theta_{\rm w}/M_{\rm W} = 1 $,
as expected.

\subsection{ Matter fields}

Up to this point we have not discussed the possible
effects derived from the introduction of matter fields,
and which will further restrict the number of allowed
theories.

In viable models it should be possible
to choose the fermion content such that it includes 
states with the \sm\ quantum numbers of the observed
quarks and leptons. In particular the choices of $\gcal $ 
and $G$ must be such that there are fermion representations
containing states with hypercharge $1/6 $.

The highest weight of a representation can be written as
\beq
\mubf_{\rm max} = \sum m_i \mubf_i = 
\sum \left( a^{-1} m \right)_j \alpbf^j \ ,
\eeq
where the $ m_i $ are integers. A generic weight in this \irrep\
is obtained by applying lowering operators $ E_{-\alpbf^i} $
associated with the simple roots. The general weight then has the form
\beq
\mubf = \sum m_j \mubf_j - \sum k_i \alpbf^i \ .
\label{eq:gen.mu}
\eeq

Now, though all the entries in the Cartan matrix
(\ref{eq:cart.mat}) $a$ are integers, this is not true
 for $ a^{-1} $; still there is an integer $N$ such that
\beq
N \left(a^{-1} \right)_{ij} = {\rm integer} \ ,
\eeq
with $N$ given in table \ref{tab:t1}
(see also appendix \ref{sec:class.groups}); also, all entries
in $ a^{-1}$ are positive.

\begin{table}[h]
$$
\begin{array}{|c|c|}
\hline
{\rm Group} & N \cr \hline
\su n & n \cr
G_2,~F_4, ~ E_8 & 1 \cr
Sp(2n), ~SO(2n+1), ~ SO(4n), E_7 & 2 \cr
E_6 & 3 \cr
SO(4n+2) & 4 \cr \hline
\end{array}
$$
\caption{Values of $N$ (see text) for
several simple Lie algebras.}
\label{tab:t1}
\end{table}

It follows that the weight of any  state is of the form
$ \mubf  = \sum (n_i/N) \alpbf^i $ where the $n_i $ are integers; then
\beq
Y \ket\mubf = (\hyp\cdot\mubf) \ket\mubf
= \left( \inv N \sum_i n_i \alpbf^i \cdot \hyp \right) \ket\mubf \ .
\eeq
In order to accommodate quarks a {\em necessary} condition is
to have
\beq
\alpbf^i \cdot \hyp = \frac N{ 6 \times {\rm integer} }
\label{eq:q.hyp}
\eeq
for at least one simple root;
which proves useful in
eliminating many groups.

The simplest way to find
sufficient conditions is to obtain $ \mubf \cdot \hyp $
using (\ref{eq:hyp.vec}) and (\ref{eq:gen.mu}),
\beq
\mubf \cdot \hyp = h \sum_{\rm isodoublets}
\left( a^{-1} m - k  \right)_{i_l},
\label{eq:exact.hyp}
\eeq
from which it can be determined by inspection whether
$ \hyp\cdot\mubf = 1/6 $ occurs in any given \irrep.

\section{Results and prospects}\label{s6}

From separately considering the various possibilities
(see appendix \ref{sec:class.groups}) for the unitary,
orthogonal and symplectic groups, as well as for the
exceptional groups $G_2,~F_4,~E_6,~E_7$, and $E_8$,
we find that the
models that can accommodate quarks, and do not
necessarily contain undesirable light $Z' $ bosons, 
are those given in the following table:
\beq
\begin{array}{|c||c|c|c|}
\hline
{\rm group} & \sw^2 & \alpbf & \hyp \cr \hline
\su{3l} & 3l/(6l-2) & \alpbf^1 & \tilde\mubf_2/2 \cr \hline
SO(2n+1) & 3/4 & \alpbf^1 & \tilde\mubf_2/6 \cr\hline
G_2 & 3/4 & \alpbf^1 & \tilde\mubf_2/6 \cr\hline
F_4 & 3/4 & \alpbf^1 & \tilde\mubf_2/6 \cr\hline
E_6 & 3/8 & \alpbf^{1,5} & \tilde\mubf_{2,3}/2 \cr\hline
E_7 & 3/4,~3/5 & \alpbf^{1,7} & \tilde\mubf_{2,3}/6 \cr\hline
E_8 & 9/16,~3/8 &\alpbf^{1,8} & \tilde\mubf_{2,3}/6 \cr\hline
\end{array}
\label{eq:models}
\eeq
where the conventions for the simple roots and weights
are given in appendix \ref{sec:class.groups}. 

Phenomenologically these models face additional obstacles.
First is the large discrepancy between the tree-level
value of $ \sw^2 $ in (\ref{eq:models})
and its low-energy experimental result of $ \sim1/4 $.
The renormalization group (RG) running between the electroweak  and
the compactification scales cannot account for this difference,
except, possibly for the $E_6$ and one $E_8$ model, in which case
$ 1/L $ would be of the same order as the GUT scale. There
is, however, also a contribution from the RG evolution
between the compactification scale and the UV cutoff $ \Lambda $
which may account for this~\cite{Dienes:1998vh,Dienes:1998vg}. 
Discussing this in detail falls outside
the scope of this paper, but will be investigated in a  future
publication.

In addition there is the remaining question of whether there
are specific compactifications for which only the desired
modes satisfy (\ref{eq:mzero}); that is, whether there is
a specific compactification choice for which $G$ is the 
full light gauge group. Again, we will not discuss this
in general but indicate the manner in which this can be realized
for one of the models considered, based on $ \gcal = \su6 $. For this case
we need to find a group $ \Gamma $ and a representation $ \vBB $
such that (\ref{eq:mzero}) has solutions only for the
zero modes associated with
the ``light'' generators  $ E_{\pm \alpbf^1},
~ \alpbf^1 \cdot \CC $ and $ \mubf_2 \cdot \CC $.

To construct $ \Gamma $ note that $\su6$ has two $\su2$
subgroups generated by
$ E_{\pm \alpbf^i}, ~\alpbf^i \cdot \CC, ~ i=3,5$  
that commute with each other and with $G$. The group
elements 
$ \exp[ (i \pi/2)  ( E_{\alpbf^i} + E_{-\alpbf^i}) ]$,
$ \exp[ (\pi/2)  ( E_{\alpbf^i} - E_{-\alpbf^i}) ]$, and
$\exp[ (i \pi/2) \alpbf^i \cdot \CC ]$, 
correspond to rotations by an angle $ \pi $ and
generate a $ \zBB_4 \times \zBB_4 $ discrete subgroup under
which only the generators of $G$ are invariant, and which
we choose as $ \Gamma $, while for $ \vBB$ we choose the
adjoint representation of these rotations~\footnote{This
argument can be extended to all models based
on $ \gcal = \su{3l} $ for $l$ even.}. Constructing the
representation $ \Rot $ then requires first the
identification of the modes that should get a \vev\
in order to have the right pattern of \ssb\ at low
energies, and then choosing the dimensionality of the
compact space that can accommodate these matrices. 
Again, this detailed calculation falls outside the scope of the present paper.

In the above considerations we have avoided discussing the difficulties
associated with Yukawa couplings, mainly because our emphasis was in obtaining
the right vector-boson and scalar sector spectrum. If all fields are bulk-propagating
these couplings are determined in terms of the gauge coupling constants and
the choice of $ \Gamma $ the representations $ \vBB $ and  $\Rot $.
Whether a model can be found that can accommodate the complicated 
fermion-mass structure without the introduction of large brane 
couplings~\footnote{Loop corrections in general induce brane couplings
\cite{Georgi:2000ks}, some of which may have power-dependence on 
the UV cutoff $ \Lambda$, but whether these affect the light
scalar modes depends on the details of the model.} 
will be investigated in a future publication.

\acknowledgements 
JW is grateful to B. Grzadkowski for various interesting comments and insights.
This work was supported by in part UC-MEXUS under Grant No. CN-08-205,  
by CONACYT and by  the U. S. Department of Energy under 
Grant No. DEFG03- 94ER40837.

\appendix

\section{Derivation of the constraints on the hypercharge generator.}

\label{sec:group.theo}

In this appendix we derive the results listed in section \ref{sec:nec.con}.

\subsection{Choice of $ \alpbf $ and the hypercharges for the simple roots}

\label{sec:ap1}

A straightforward examination of the roots of all simple groups shows that
{\em any} root $ \alpbf $  can be transformed into one of the simple roots
by appropriate permutation and inversion of the axes. It follows that there
is an orthogonal matrix $ \oBB $ such that
\beq
\oBB \alpbf = \sa \ .
\eeq
for some simple root $\sa$.
Now, starting from the original set of simple roots 
$ \{ \alpbf^i \}$ we define a new set by
\beq
\alpbf^i_{\rm new} = \oBB^T \alpbf^i \ ,
\eeq
which also constitutes a set of simple roots since they generate the right
Cartan matrix~\cite{Jacobson:1979}. In particular $ \sa_{\rm new} = \alpbf $, so
we can indeed assume $ \alpbf $ is a simple root. Using (\ref{eq:def.of.s.sz})
and $ \alpbf^i \cdot \alpbf^j \le 0 ~ (i \not= j)$, it follows that for simple
roots (except $ \alpbf$ itself), $ s = - s_z $.

A positive root is defined as one whose first non-zero component
is positive; all simple roots are positive. Using $ \hat\hyp $ as the
first coordinate unit vector then implies that we can choose all
simple roots to have non-negative hypercharge: $ \hyp \cdot\alpbf^i
\ge 0 $

It now proves convenient to divide the simple roots into 3 sets
(some of which may be empty):
\bit
\item $\alpbf^{i_k}$,  with non-negative hypercharge: 
$\hyp\cdot \alpbf^{i_k}  = h^k > 0 $.
\item $ \gambf^r$,  with zero hypercharge and non-zero isospin
(like $ \alpbf$): $ \hyp \cdot\gambf^r =0,~ \alpbf\cdot\gambf^r \not=0 $
(in fact, for $ \gambf^r \not= \alpbf $ one has $ \alpbf \cdot \gambf^r < 0 $,
since $\alpbf$ and $ \gambf^r $ are simple roots).
\item $\zetbf^l$, which are $G$-singlets: $ \hyp \cdot\zetbf^l = \alpbf\cdot\zetbf^l =0$.
\eit
Using the rescaled fundamental weights $ \tilde\mubf_i $ defined in
(\ref{eq:def.of.mu}) we can then write
\beq
\hyp = \sum h^k \tilde\mubf_{i_k} \then
\hyp\cdot\alpbf^{i_k} = h^k,~~ \hyp\cdot\gambf^r = \hyp\cdot\zetbf^l =0 \ .
\eeq

\subsection{Orthogonality of the $ \zetbf^l$ and $ \hyp'_q $}
\label{sec:ap2}

If there is at least one $ \zetbf^l $,
the set of generators $ \{ E_{\zetbf^l} ,~ \CC \cdot\zetbf^l \} $
together with all their commutators
form a sub-algebra $S_0 $ that commutes with $G$.

Denote by $M_q$ the collection of all roots that have a specific set of
$G$ quantum numbers, $ q = \{ h, s, s_z\}$. Then, since all
elements of $S_0 $ are $G$ singlets,
$ [ S_0 , M_q] \subset M_q $. The elements of $M_q$ then carry
a representation of $S_0 $ (in general reducible). Now consider
the trace of $ \vv\cdot \CC $, with $ \vv\cdot\alpbf =0 $,
in this representation. We denote the states by
$ \ket\betbf $ with $ E_\betbf \in M_q $, which,
using~(\ref{eq:generatoralgebra}), obey
\beq
 C_i \ket\betbf = \beta_i \ket\betbf
\eeq
and, using (\ref{eq:norm}),
$ \vevof{\betbf | \gambf } = {\rm tr} E_{-\betbf} E_\gambf
= \delta_{\betbf, \gambf} $. Then
\beq
{\rm tr}_{M_q} \left\{ \vv\cdot\CC \right\}
=  \sum_{\betbf \in M_q} \vevof{ \betbf | \vv\cdot\CC | \betbf}
=  \sum_{\betbf \in M_q} \vv\cdot\betbf
= \vv\cdot \hyp'_q \ .
\label{eq:v.ypq}
\eeq
Now we decompose $M_q $ into \irrep s of $S_0 $: $ M_q = \oplus M_q\up r $.
Any weight $ \mubf $ in any of the \irrep s
has the useful property that its Weyl-reflection is also a weight
in that same \irrep:
\beq
\mubf: ~{\rm weight}
\then \mubf' =
\mubf - \left(\frac{2 \mubf \cdot \zetbf^l}{|\zetbf^l|^2}
\right) \zetbf^l: ~
{\rm weight} \quad \forall \zetbf^l \in S_0 \ ;
\eeq
in particular,
$ \mubf\cdot \zetbf^l = - \mubf' \cdot \zetbf^l $.
Using (\ref{eq:v.ypq}) this implies that, for all $ q \not=0 $,
\beq
{\rm tr}_{M_q} \left\{\zetbf^l \cdot \CC \right\}
 = \sum_r {\rm tr}_{M_q\up r}\left\{ \zetbf^l \cdot \CC \right\}
= 0 \then \zetbf^l \cdot \hyp'_q =0 \ ,
\label{eq:zeta.ypq}
\eeq
since the diagonal elements cancel in pairs because of
the above Weyl-reflection property. In particular, the last term in
(\ref{eq:vbm}) is zero.

Now, if $q$ corresponds to zero hypercharge then both $ E_\betbf $
{\em and} $E_{- \betbf} $ will be members of $M_q $
(because $\su2$ has no complex
representations), and the corresponding $\hyp'$ vanishes.

\subsection{General form and some properties of the $\hyp'_q $.}
\label{sec:ap3}

Any positive root $ \betbf $ can be written as a linear
combination of the simple roots with non-negative integer coefficients.
It follows that the summations in (\ref{eq:sum.1}) and
(\ref{eq:sum.2}), for positive
hypercharge roots, have the same property; whence
\beq
\hyp'_q =  \sumprime_{}\betbf_\perp =
\sum_k \alpbf^{i_k}_\perp  n_k + \sum_r \gambf^r_\perp  n_r +
\sum_l \zetbf^l  n_l \ , \quad n_{k,r,l} \ge0
\label{eq:ypq.s.r}
\eeq
where, as before, 
$ \betbf_\perp = \betbf - (\hat\alpbf \cdot \betbf) \hat\alpbf$,
so that
$ \zetbf^l_\perp = \zetbf^l $. For this to be orthogonal to the $ \zetbf^l $ as required by
(\ref{eq:zeta.ypq}) we must have
\beq
\sum_k a^{l i_k}  n_k + \sum_r a^{lr}  n_r +
\sum_l a_0^{ll'}  n_{l'} =0
\then
n_l = - \sum_k \left( a_0^{-1} a \right)_l{}^{i_k}  n_k
- \sum_r \left( a_0^{-1} a \right)_l{}^r  n_r \ ,
\label{eq:nl}
\eeq
where $a$ is the Cartan matrix (\ref{eq:cart.mat}) of the full algebra,
and $a_0 $ that of $S_0$.

Assume that not all the $ h^k $ are equal; then there
are at least two distinct values and we can separate
$ h^k = h_a,~ k \in K_a, ~ a=1,2 $ (there may, of course,
be other values); then there
will be at least two $ \hyp'_q $ not proportional to one another.
To see this consider
the $ \hyp'_{q_a} $ with $q_a$ the $G$ quantum numbers of $ \alpbf^{i_k},
~k\in K_a $. Then the corresponding $ \betbf $
in (\ref{eq:ypq.s.r}) must have $ n_r =0 $
since $ \gambf^r $ carry non-zero isospin; using then (\ref{eq:nl}),
\beq
\hyp'_{q_a} = \sum_{k \in K_a} n_k \left\{ \alpbf^{i_k}_\perp -
\sum_l \zetbf^l \left( a_0^{-1} a \right)_l{}^{i_k} \right\} \ .
\label{eq:simple.yp}
\eeq
From this, if $k \in K_1$, then
$ \tilde\mubf_{i_k} \cdot \hyp'_{q_1} \not=0 $, but
$ \tilde\mubf_{i_k} \cdot \hyp'_{q_2} =0$ since 
$K_1 \cap K_2 = \emptyset $
(and similarly for $ k \in K_2 $), so that
$ \hyp'_{q_1} $ and $ \hyp'_{q_2} $ cannot be parallel.
Equivalently,
\beq
\hyp_{q_1} \propto \hyp_{q_2} \then h_1 = h_2 .
\eeq
If all the $ h^k $ are the same but the corresponding
$ \alpbf^{i_k} $ have different isospin, we can repeat
the above argument and show that there will again be two
$ \hyp' $ not proportional to one another.

A necessary condition for the absence of
additional light vector bosons is then for all the $h^k$
to be equal, and for the roots $ \alpbf^{i_k} $
to have the same isospin. Since $ \rho = 1 $ at tree level requires
that only isodoublets acquire a \vev,
we obtain (\ref{eq:hyp.vec}).

The absence of roots of the type $ \gambf^r $ (aside from $\alpbf $)
follows from the requirement that $\hyp$ be parallel to
all the $ \hyp'_q $, that is, $ \lambda \hyp = \hyp'$.
Taking $ \hyp'$ of the form  (\ref{eq:simple.yp}),
using (\ref{eq:hyp.vec}), and 
dotting with $ \tilde\mubf_r $ (dual to $ \gambf^r 
\not= \alpbf$), we find
\beq
\lambda h \sum_k \frac2{|\alpbf^{i_k} |^2} a^{-1}_{r\,i_k}= 0 \ ,
\eeq
where we used the fact that $ a_0$ has non-vanishing elements only in
the $ \zetbf$ subspace (so that, for example,  $\left(
a_0^{-1} a \right)_r{}^{i_k} =0 $); since
$ a^{-1} $ has only positive elements this
equation is impossible to satisfy {\em unless} there are no
roots of the type $ \gambf $.

\section{The classical groups}
\label{sec:class.groups}

In this appendix we provide some relevant details for most of
the classical groups (absent are $ E_{7,8} $)~\cite{Gilmore:1974,Slansky:1980mb,Slansky:1981yr,Georgi:1982jb}. 
In discussing
the various possible choices of $ \alpbf $ and \hyp\ 
we will follow the following steps:
\ben
\item We first obtain the simple roots, the fundamental weights,
the Cartan matrix and its inverse; from this last we obtain the 
values of $N$ in table \ref{tab:t1} that are used in (\ref{eq:q.hyp}).
\item For each choice $ \alpbf = \alpbf^i $ ($\alpbf^i $ denote the simple roots)
we find the isospin of all the remaining roots. If some roots
carry isospin $ > 1/2 $ then we are guaranteed to have additional light
$Z' $ and the model is discarded.
\item For groups whose
simple roots carry isospin $0$ and $1/2$ we construct the hypercharge
vector \hyp\ using (\ref{eq:hyp.vec}); we then find all the $ \hyp'_q $
and discard the group if any one of these vectors is not parallel to \hyp.
We repeat this for each choice of $ \alpbf $.
\item Finally we discard groups inconsistent with (\ref{eq:q.hyp}),
while for those that do satisfy this equation we calculate
explicitly the hypercharges using (\ref{eq:exact.hyp}), 
and discard those groups which cannot accommodate quarks.
\een
In the following we shall frequently use the notation $ \betbf>0\uparrow$ to
denote positive isodoublet roots with $ s_z = +1/2 $.

\subsection{$A_n = \su{n+1}$}
\setcounter{paragraph}{0}

\paragraph{\color{blue}{Roots and weights}}
For this series, the roots are $ \ee^k - \ee^j, ~ 1\le k \not= j \le n+1 $,
the positive roots $\ee^k - \ee^j , ~ 1\le k < j \le n+1 $ while the
simple roots are
\beq
\alpbf^k = \ee^k - \ee^{k+1} ,~~ 1 \le k \le n \ ,
\eeq
where $ (\ee^j)_i = \delta_{ij}$.
Note that the space involved has $ n+1 $ dimensions, while there
are only $n$ Cartan generators, this is ``fixed'' by requiring
all vectors to be orthogonal to
\beq
\wp = (1,1, \cdots , 1) \ .
\eeq

The fundamental weights are
\beq
\tilde\mubf_k = \mubf_k = \ff^k - \frac k{n+1} \wp\,; \quad \ff^k = \sum_{i=1}^k \ee^i \ .
\label{eq:def.of.f}
\eeq

\paragraph{\color{blue}{Cartan matrix}}
From the definition (\ref{eq:cart.mat})
\beq
a(A_n) = \bpm{2 & -1 & 0 & \cdots &0 \cr
-1 & 2 & -1 & \cdots & 0 \cr
0 & -1 & 2 & \cdots & 0 \cr
\vdots & \vdots & \vdots & \ddots & \vdots \cr
0 & 0 & 0 & \cdots & 2 }  \ , \epm
\eeq
which is symmetric. Taking the inverse,
\beq
(n+1) \left( a^{-1} \right)_{ij} = \hbox{integer}
\then N = n+1 \ .
\eeq

\paragraph{\color{blue}{$\su2$ content}}

A direct inspection of the
Cartan matrix shows that the ratio of an off-diagonal element to a diagonal
element is either $0$ or $ -1/2 $, which implies that all simple
roots re isosinglets or isodoublets; alternatively, the same results
follows from the fact that
for any two different roots $ \betbf \cdot \gambf/\betbf \cdot\betbf = 0,-1$. 
For $ \alpbf = \alpbf^k $
\beq
\betbf>0\uparrow: \quad \ee^k - \ee^j,~ \ee^l - \ee^{k+1}, \quad
l < k,~ j>k+1
\eeq
(for $ k = 1 $ there are no roots with $ l< k = 1 $). As mentioned above, this implies that
it is possible to satisfy the $\rho=1$ constraint with this group. This result applies to all simple
groups as can be seen from their Cartan matrices.

\paragraph{\color{blue}{$\ui$ content}}

For $ \alpbf = \alpbf^k $ the only simple roots that  carry non-zero
isospin are $ \alpbf^{k \pm 1 } $, it follows from (\ref{eq:hyp.vec}) that
\beq
\hyp = h \left( \tilde\mubf_{k-1} + \tilde\mubf_{k+1} \right) \ ,
\eeq
with the convention that $ \mubf_{k=0} = 0 $;
(\ref{eq:sz.and.h}) then implies $h=1/2 $.
The only non-trivial $ \hyp' $ is given by
\beq
 \hyp'_{h,1/2,1/2} =
\frac{n-2k+3}2 \mubf_{k+1} - \frac{n-2k-1}2 \mubf_{k-1} \ ,
\eeq
which, but for two special cases, is not parallel to \hyp, so there will
be undesirable light $ Z' $. The two exceptions for which $ \hyp' \propto \hyp$ are:
\bit
\item $ k = 1 $:
\beq
\alpbf = \alpbf^1 \,, \quad
\hyp =\half\mubf_2\,, \quad
\sw^2 = \frac{n+1}{2n} \ .
\eeq

\item $n$ odd and $ k = (n+1)/2 $:
\beq
\hyp = \half \left( \mubf_{(n+3)/2} + \mubf_{(n-1)/2} \right)\,; \qquad
\hat\alpbf = \alpbf^{(n+1)/2}\,;\qquad
 \sw^2 = \frac2{n+1} \ .
\eeq

\eit

\paragraph{\color{blue}{Matter content}}

Since $ \alpbf^i \cdot\hyp = \pm1/2,~0 $, we need only require
$N=n+1$ to be a multiple of $3$ for (\ref{eq:q.hyp}) to be met.
However, for the case where  $ k = (n+1)/2 $ and $n$ is odd,
$ 2 \hyp\cdot\mu_k $ is an integer, which implies that the
highest weight of all multiplets has half-integer hypercharge;
and the raising and lowering operators can only change
the hypercharge by $1/2 $. This implies that, in fact, this
case cannot accommodate quarks.

In contrast, models with $ \alpbf = \alpbf^1,
~ \hyp = \mubf_2/2 $ satisfy
(\ref{eq:q.hyp}) provided $N = 3 l $ for an integer $l$.
In this case
(\ref{eq:hyp.vec}) and (\ref{eq:gen.mu}) give
\bea
h = \hyp \cdot \mubf
& = &\sum_{j=1}^{3l-1} m_j \left( 1 - \frac j{3l} \right) - \half k_2 \cr
s_z =\frac{\alpbf \cdot \mubf}{|\alpbf|^2}
&=& \half m_1 - k_1 + \half k_2 \ ,
\eea
where $k_1,~ k_2 $ and the $ m_k $ are non-negative integers,
so that there
{\em will} be states with hypercharge $1/6 $. The simplest case
corresponds to $ l = 2 $, $\su6 $; for example, the state with highest
weight $ \mubf_4 $ (a {\bf15} of $\su6$) decomposes into
\beq
\begin{array}{|c|c|}
\hline
(h,s_z) & {\rm multiplicity} \cr \hline
(1/3,0) & 6 \cr \hline
(-2/3,0) & 1 \cr \hline
(-1/6,1/2) & 4 \cr \hline
(-1/6,-1/2) & 4 \cr \hline
\end{array}
\eeq

\subsection{$ SO(2n+1) = B_n $}
\setcounter{paragraph}{0}

\paragraph{\color{blue}{Roots and weights}}
The roots are $\ee^i \pm
\ee^j, ~i\not=j$ and $ \pm \ee^i$; the positive roots are $\ee^i \pm
\ee^j, ~i < j$ and $ \ee^i$; and the simple roots are
\bea
\alpbf^k &=& \ee^k-\ee^{k+1},\quad k\le n-1 \cr \alpbf^n &=&
\ee^n.
\eea
The corresponding fundamental weights are (see eq. \ref{eq:def.of.f})
\beq
\mubf_k = \ff^k,~~ k \le n-1\,; \quad \mubf_n = \half \ff^n
\then \tilde\mubf_k = \ff^k, ~~ k \le n.
\eeq

\paragraph{\color{blue}{Cartan matrix}}
From the definition
\beq
a(B_n) = \bpm{
 2 & -1 &  0 & \cdots &  0 &  0 &  0 \cr
-1 &  2 & -1 & \cdots &  0 &  0 &  0 \cr
 0 & -1 &  2 & \cdots &  0 &  0 &  0 \cr
\vdots & \vdots & \vdots & \ddots & \vdots& \vdots& \vdots \cr
 0 &  0 &  0 & \cdots &  2 & -1 &  0 \cr
 0 &  0 &  0 & \cdots & -1 &  2 & -2 \cr
 0 &  0 &  0 & \cdots &  0 & -1 &  2 \cr
} \ , \epm
\eeq
which is not symmetric. Also, taking the inverse
\beq
2 \left( a^{-1} \right)_{ij} = \hbox{integer}
\then N = 2 \ .
\eeq

\paragraph{\color{blue}{$\su2$ content}}
When $ \alpbf =\alpbf^n$ there are no isodoublet roots, so we will not
consider it further. If $ \alpbf = \alpbf^k, ~ k<n$, all roots are
either isosinglets or isodoublets; for the latter
\beq
\betbf>0\uparrow: \quad
\ee^k \pm \ee^j,~ \ee^l + \ee^k,~ \ee^l - \ee^{k+1},~ \ee^k,\quad l<k,~ j>k+1 \ .
\eeq

\paragraph{\color{blue}{$\ui$ content}}
If $ \alpbf = \alpbf^k ~ (k<n)$, the only isodoublet simple roots
are $ \alpbf^{k \pm1 }$, hence, using (\ref{eq:hyp.vec}),
\beq
\hyp = h \left( \ff^{k+1} + \ff^{k-1} \right), \quad 1 \le k < n
\eeq
(again with the convention $ \ff^0 = 0 $).
Then roots can have hypercharge $h$ or $3h$,
\beq
h:~\{ \ee^k \pm \ee^j,~ \ee^l - \ee^{k+1}, ~ \ee^k  \}; \quad
3h:~ \{ \ee^l + \ee^k \};\quad l<k,~ j>k+1
\eeq
so that (\ref{eq:sz.and.h}) requires $ h = 1/2$ or $ h=1/6 $.

As for the $ \hyp' $, using the $ \betbf>0\uparrow$,
\bea
\hyp'_{h,1/2,1/2}&=& \frac{2n-3k}2 \ff^{k+1}+ \frac{2 - 2n + 3k}2 \ff^{k-1} \cr && \cr
\hyp'_{3h,1/2,1/2} &=&\frac{k-1}2 \ff^{k+1} + \frac{3-k}2\ff^{k-1} \ .
\eea
These are not parallel to $\hyp$ {\em except}~\footnote{
The case where the
coefficients of $ \ff^{k \pm 1} $ in $\hyp'_q$  are equal corresponds to
$ n-3k/2 = 1 - n + 3k/2 $ {\em and} $ k-1=3-k $, which have no solutions
when $ k$ and $n$ are integers.}
 when $ k = 1 $, in which case $ \hyp = h \ff^2 = h ( \ee^1 + \ee^2 ) $.
Then
\beq
\alpbf = \alpbf^1\,, \quad
\hyp = \half \tilde\mubf_2\,, \quad \sw^2 = \half\,; \qquad
{\rm or} \qquad
\alpbf = \alpbf^1\,, \quad
\hyp = \inv6 \tilde\mubf_2\,, \quad \sw^2 = \frac34\, .
\eeq

\paragraph{\color{blue}{Matter content}}

Since $N=2,~ \alpbf^i \cdot \hyp = 0,h $,
models with $ h=1/2 $ cannot accommodate quarks. But
models with $ h=1/6 $ can. For example, for
$ n = 2 $, the $ SO(5) $ the multiplet with highest
weight $ \mubf_2 $ has dimension 4 and contains an
isodoublet of hypercharge $1/6 $.

\subsection{$Sp(2n) = C_n$ }
\setcounter{paragraph}{0}

\paragraph{\color{blue}{Simple roots}}

For this group the roots are  $ \pm \ee^i \pm \ee^j $ and
$ \pm 2 \ee^i $, the positive roots are $ \ee^i \pm \ee^j,~i<j$ and $
2\ee^i $, and the simple roots are
\beq
\alpbf^k = \ee^k - \ee^{k+1}, ~ 1 \le k \le n-1; \quad \alpbf^n =
2\ee^n \ .
\eeq
Then
\beq
\mubf_k = \tilde\mubf_k = \ff^k , ~ k < n\,; \quad
\mubf_n = 2 \tilde\mubf_n = \ff^n \ .
\eeq

\paragraph{\color{blue}{Cartan matrix}}
For this group the Cartan matrix is the transpose of the
one for $B_n $, hence $ N = 2 $.

\paragraph{\color{blue}{$\su2$ content}}

If $ \alpbf = \alpbf^k, ~ k<n$ the roots $ 2\ee^k $ have $s=1 $
so this case need not be considered further.
If $ \alpbf = \alpbf^n $, then all roots are isodoublets or
isosinglets; in particular,
\beq
\betbf>0\uparrow: \quad \ee^l + \ee^n,\quad l<n \ .
\eeq

\paragraph{\color{blue}{$\ui$ content}}
For $ \alpbf = \alpbf^n $ only $ \alpbf^{n-1} $ carries isospin, hence
(\ref{eq:hyp.vec}) yields
\beq
\hyp = h \tilde\mubf_{n-1} = \half h \ff^{n-1} \ ,
\eeq
and all the $\betbf>0\uparrow $ roots have hypercharge $h/2$; hence we
choose $ h = 1 $. As for the $\hyp' $ we only have
\beq
\hyp'_{h/2,1/2,1/2} = \ff^{n-1} \ ,
\eeq
so there are no additional $ Z' $. Using then (\ref{eq:tw})
\beq
\alpbf = \alpbf^n \,, \quad
\hyp = \half \mubf_{n-1}\,,\quad
\sw^2 = \inv n \ .
\eeq

\paragraph{\color{blue}{Matter content}}

For $ \alpbf = \alpbf^n $  we get $ \hyp \cdot\alpbf^i = 0,~1/2$,
so (\ref{eq:q.hyp}) cannot be met since $ N = 2$.

\subsection{$ SO(2n) = D_n $}
\setcounter{paragraph}{0}

\paragraph{\color{blue}{Simple roots}}
The roots are $ \pm \ee^i
\pm \ee^j,~ i\not=j$, the positive roots are $ \ee^i \pm \ee^j,~ i <
j$, and the simple roots are
\bea
\alpbf^k &=& \ee^k-\ee^{k+1},~\quad k = 1, \ldots n-1 \cr
\alpbf^n &=& \ee^{n-1}+\ee^n \ .
\eea

The fundamental weights are (see eq. \ref{eq:def.of.f})
\beq
\mubf_k = \ff^k,~~ k<n-1\,; \quad
\mubf_{n-1} = \ff^{n-1} - \half \ff^n \,; \quad
\mubf_n = \half \ff^n \ ,
\eeq
with $ \tilde\mubf_k = \mubf_k $ in all cases.

\paragraph{\color{blue}{Cartan matrix}} From the definition
we find
\bea
a(D_n) &=& \bpm{2 & -1 & 0 & \cdots & 0 & 0 & 0 \cr
-1 & 2 & -1 & \cdots & 0 & 0 & 0  \cr
0 & -1 & 2 & \cdots & 0  & 0 & 0 \cr
\vdots & \vdots & \vdots & \ddots & \vdots& \vdots& \vdots \cr
0 & 0 & 0 & \cdots & 2 & -1 & -1\cr
0 & 0 & 0 & \cdots & -1 & 2 & 0\cr
0 & 0 & 0 & \cdots & -1 & 0 & 2\cr
} \epm
\eea
which is symmetric. Taking the inverse we find
\beq
\left[3 + (-1)^{n+1} \right] \left(a^{-1}\right)_{ij} = \hbox{integer}
\then N = \left\{
\begin{array}{ll}
4 & \hbox{for}~n~\hbox{odd} \cr
2 & \hbox{for}~n~\hbox{even}
\end{array} \right.
\eeq

\paragraph{\color{blue}{$\su2$ content}}
For any choice $ \alpbf = \alpbf^k $
all the roots are isodoublets or isosinglets, in particular
the  $ \betbf>0\uparrow $ roots are
\bea
\alpbf = \alpbf^k, ~ k<n: &&
\ee^i- \ee^{k+1},~ \ee^i + \ee^k ~~ i< k\,;
\cr &&
\ee^k \pm \ee^l,~~ l> k+1 \cr && \cr
\alpbf = \alpbf^n && \ee^i + \ee^{n-1},~ i<n-1\,;
\quad  \ee^i+ \ee^n,~ i<n \ .
\eea

\paragraph{\color{blue}{$\ui$ content}}

For
$ \alpbf = \alpbf^k,~ k<n-2 $ the only simple
isodoublet roots are $ \alpbf^{k\pm1}$; for
$ \alpbf = \alpbf^{n-2} $ the simple
isodoublet roots are $ \alpbf^{n,~ n-1,~ n-3}$; for
$ \alpbf = \alpbf^{n-1, ~n}  $ the only simple
isodoublet root  $ \alpbf^{n-2}$.
Still when constructing \hyp\
these reduce to only two cases:
\bea
\alpbf = \alpbf^k,~ k \le n-2: && \hyp = h \left(
\ff^{k+1} + \ff^{k-1} \right) \cr
\alpbf = \alpbf^k,~ k \ge n-1: && \hyp = h \ff^{n-2} \ .
\eea

All the $ \betbf>0\uparrow $ have hypercharge $h$,
except
$ \ee^i + \ee^k$ and $i < k < n-1 $ which have hypercharge
$3h$. Then
\bit
\item $ \alpbf = \alpbf^k,~ k< n-1$:
\bea
\hyp'_{h,1/2,1/2} &=&
\frac{2n - 3k - 1}2 \ff^{k+1} + \frac{3k - 2n + 3}2 \ff^{k-1} \cr && \cr
\hyp'_{3h,1/2,1/2} &=& \frac{k-1}2 \ff^{k-1} + \frac{3-k}2 \ff^{k+1} \ .
\eea

Of these, the only case~\footnote{The other possibility is
to choose value of $n$ and $k$ so that the coefficients
of $ \ff^{k\pm1} $ are equal, that is
$ k-1 = 3-k $ and $ n-3k/2-1/2 = 3k/2 - n +3/2 $;
but these have no integer solutions.}
where there are no $Z'$ corresponds to $ k = 1 $, in this
instance $ \hyp = h \ff^2 = h ( \ee^1 + \ee^2 ) $
while (\ref{eq:sz.and.h}) requires $ h = 1/2$; then we have
\beq
\alpbf = \alpbf^1 \,,\quad
\hyp = \half \mubf_2 \,,\quad
\sw^2 = \half \ .
\eeq

\item $ \alpbf = \alpbf^{n-1},~\alpbf^n $, then $ \hyp = h \ff^{n-2}$
and
\beq
\hyp'_{h,1/2,1/2} = \ff^{n-2} \, . \qquad
\eeq
Also, (\ref{eq:sz.and.h}) requires $ h =1/2 $ so that
\beq
\alpbf = \alpbf^n,~\alpbf^{n-1}\,,\quad
\hyp = \half \mubf_{n-2} \,,\quad
\sw^2 = \frac2n \ .
\eeq

\eit

\paragraph{\color{blue}{Matter content}}

Since $ N=4,2$ and $ \hyp \cdot \alpbf^i=0,1/2$
such models cannot accommodate quarks.

\subsection{$F_4$}
\setcounter{paragraph}{0}

\paragraph{\color{blue}{Simple roots}}
The roots are
\bea
\pm \left( \ee^k + \ee^j \right), \quad \pm \left( \ee^k - \ee^j
\right), \quad\pm \ee^k, ~ (1 \le k < j \le 4); \qquad \pm\half
\sum_{k=1}^4\left( \pm \ee^k \right) \ .
\eea
The simple roots are
\bea
\alpbf^1 = \ee^2 - \ee^3 \quad && \quad \alpbf^2 = \ee^3 - \ee^4\cr
\alpbf^3 = \ee^4 \quad && \quad \alpbf^4 = \half \left( \ee^1 - \ee^2
- \ee^3 - \ee^4 \right) \ .
\eea

The fundamental weights are then
\bea
\mubf_1 = \tilde\mubf_1 = \ee^1 + \ee^2   \ , \quad && \quad
\mubf_2 = \tilde\mubf_2 = \ee^1 + \ff^3  \ , \cr
\mubf_3 = \half \tilde\mubf_3 = \ee^1 + \half \ff^4   \ , \quad && \quad 
\mubf_4 = \half \tilde\mubf_4 = \ee^1  \ .
\eea

\paragraph{\color{blue}{Cartan matrix}}
From the above
\beq
a(F_4) = \bpm{
 2 & -1 &  0 &  0 \cr
-1 &  2 & -1 &  0 \cr
 0 & -2 &  2 & -1 \cr
 0 &  0 & -1 &  2 }\epm
\qquad
a(F_4)^{-1} = \bpm{
2 & 3 & 2 & 1 \cr
3 & 6 & 4 & 2 \cr
4 & 8 & 6 & 3 \cr
2 & 4 & 3 & 2 } \ , \epm
\eeq

so that $ N=1 $.

\paragraph{\color{blue}{$\su2$ content}}

A simple Mathematica program shows that $\alpbf = \alpbf^{3,4}$
generate isotriplets and isodoublets, so we will not consider them
further. The other possibilities have only isodoublets (or isosinglets).

\paragraph{\color{blue}{$\ui$ content}}

For $ \alpbf = \alpbf^2 $ we have
$ \hyp = h ( \tilde\mubf_1 + \tilde\mubf_3 ) $; then
the isodoublets can have hypercharge $h,~3h$ or $5h$, while the
isosinglets have hypercharge $ 2h,~4h$ or $6h$. A straightforward
calculation then gives
\bea
2\hyp'_{5h,1/2,1/2}  = \hyp'_{2h,0,0}&=& (2,0,1,1) \ , \cr
\hyp'_{4h,0,0}&=&\half(3,1,1,1) \ , \cr
2\hyp'_{h,1/2,1/2}  = \frac23 \hyp'_{3h,1/2,1/2}  = \hyp'_{6h,0,0}&=&(1,1,0,0) \ ,
\eea
all of which are linear combinations of \hyp\ and
$(1,-5,3,3) $, so there will be at least one light $Z'$.

The remaining possibility is $ \alpbf = \alpbf^1 $
for which $ \hyp = h \tilde\mubf_2 $; then
the isodoublets can have hypercharge $h$ or $3h$ while the
isosinglets have hypercharge $ 2h $. A straightforward
calculation then gives
\beq
\hyp'_{h,1/2,1/2} = 2 \hyp'_{3h,1/2,1/2}
= \half \hyp'_{2h,0,0} =  (2,1,1,0) \ ,
\eeq
so there is no undesirable $Z'$ and
(\ref{eq:sz.and.h}) require $ h=1/2,~1/6 $; then
\beq
\alpbf = \alpbf^1 \,,\quad
\hyp = \half \mubf_2 \,,\quad
\sw^2 = \inv4 \,; \qquad {\rm or} \qquad
\alpbf = \alpbf^1 \,,\quad
\hyp = \inv6 \mubf_2 \,,\quad
\sw^2 = \frac34 \ .
\eeq

\paragraph{\color{blue}{Matter content}}
We find a result similar to the case $B_n$:
(\ref{eq:q.hyp}) can be met only for $ h=1/6 $ .
Using (\ref{eq:hyp.vec}) and (\ref{eq:gen.mu}) we find
that for this choice of $h$
the state with weight $ \mubf$ has the following $G$ quantum numbers:
\bea
h &=& \inv6 \left( 3 m_1 + 6 m_2 + 4 m_3 + 2 m_4 - k_2 \right) \ , \cr
s_z &=& \half \left(m_1 - k_2 \right) - k_1 \ ,
\eea
so there are multiplets with hypercharge $1/6 $. For example,
the {\bf 26} decomposes into
\beq
\begin{array}{|c|c|}
\hline
(h,s_z) & {\rm multiplicity} \cr \hline
(\pm1/3,0) & 1 \cr \hline
(\pm2/3,0) & 2 \cr \hline
(\pm1/6,\pm1/2) & 2 \cr \hline
(\pm5/6,\pm1/2) & 1 \cr \hline
(\pm1/2,\pm1/2) & 1 \cr \hline
(0,\pm1) & 1 \cr \hline
(0,0) & 2 \cr \hline
\end{array}
\eeq

\subsection{$G_2$}
\setcounter{paragraph}{0}

\paragraph{\color{blue}{Simple roots}}
The positive roots are
$ (\ee^1 \pm \sqrt{3} \ee^2)/2 $,
$ (\ee^1 \pm \ee^2/\sqrt{3})/2 $,
$\ee^1$ and $ \ee^2/\sqrt{3} $;
 the simple roots are
\beq
\alpbf^1 = \inv{\sqrt{3}}\ee^2, \qquad
\alpbf^2 = \half \left( \ee^1 - \sqrt{3} \ee^2 \right) \ ,
\eeq
and the fundamental weights are
\beq
\mubf_1  = \inv6 \tilde\mubf_1
= \inv{2\sqrt{3}}\left( \sqrt{3}\ee^1 + \ee^2 \right), \qquad
\mubf_2 = \tilde\mubf_2 = 2 \ee^1 \ .
\eeq

\paragraph{\color{blue}{Cartan matrix}}
 We find
\beq
a(G_2) = \bpm{ 2 & -3 \cr -1 & 2 }\epm, \quad
a(G_2)^{-1} = \bpm{ 2 & 3 \cr 1 & 2 }\epm,
\eeq
so that $N=1 $.

\paragraph{\color{blue}{$\su2$ content}}

For $ \alpbf = \alpbf^1$ all roots are either isosinglets
or carry isospin $3/2 $, so we will not consider this case further.
For $ \alpbf = \alpbf^2 $ there are only isodoublets
and isosinglets; explicitly $ \betbf>0\uparrow:
~(\ee^1-\ee^2/\sqrt{3})/2$ and $\ee^1 $.

\paragraph{\color{blue}{$\ui$ content}}
For $ \alpbf = \alpbf^2 $ we have $ \hyp = h \tilde\mubf_1 $
and all $\betbf>0\uparrow$ roots have hypercharge
$h$ or $3h$, while the isosinglets have hypercharge $2h$.
Using this we find
\beq
12\hyp'_{h,1/2,1/2}  = 4 \hyp'_{3h,1/2,1/2}
= 6\hyp'_{2h,0,0} = \tilde\mubf_1\ ;
\eeq
(of course, there being only two generators, there cannot be additional $Z'$),
while (\ref{eq:sz.and.h}) requires $ h=1/2$ or $h=1/6 $; hence
\beq
\alpbf = \alpbf^2 \,,\quad
\hyp = 3 \mubf_1 \,,\quad
\sw^2 = \inv4 \,; \quad {\rm or} \quad
\alpbf = \alpbf^2 \,,\quad
\hyp = \mubf_1 \,,\quad
\sw^2 = \frac34 \ .
\eeq

\paragraph{\color{blue}{Matter content}}

For $ \hyp = h\tilde\mubf_1 $,
$ \hyp \cdot \alpbf^i = 0,~h$, so that only the case $ h=1/6 $
can satisfy (\ref{eq:q.hyp}), since $ \hyp\cdot \mubf_i =
1/3,1 $ there will be states with hypercharge $1/6 $.
For example, the multiplet with highest weight
$ \mubf_1 $ (a {\bf14} of $G_2$) contains an isosinglet
of hypercharge $1/3$ and an isodoublet of hypercharge $1/6 $.

\bigskip

The possibility of using $\gcal = G_2$  was extensively studied in
Ref. \cite{Csaki:2002ur} for the case $ h=1/2 $, including the
possibility of overcoming the quark hypercharge difficulties
through appropriate brane couplings.

\subsection{$E_6$}
\setcounter{paragraph}{0}

\paragraph{\color{blue}{Simple roots}} The roots are
$ \pm  \ee^k \pm \ee^j, ~ 1 \le k < j \le 5$ (40 vectors)
and $\left(\sum_{k=1}^5 \xi_j \ee^k + \sqrt{3}\xi_6 \ee^6 \right)/2$ with
$ \xi_k^2=1,~\prod_{k=1}^6\xi_k= + 1 $ (32 vectors). The simple roots are
\bea
\alpbf^1 = -\half (1, 1, 1, 1, 1, \sqrt{3}), &\qquad&
\alpbf^2 = (1, 1, 0, 0, 0, 0), \cr
\alpbf^3 = (0, -1, 1, 0, 0, 0), &\qquad&
\alpbf^4 = (0, 0, -1, 1, 0, 0), \cr
\alpbf^5 = (0, 0, 0, -1, 1, 0), &\qquad&
\alpbf^6 = (-1, 1, 0, 0, 0, 0) \ .
\eea

The fundamental weights in this basis are
\bea
\mubf_1 = (0, 0, 0, 0, 0, -2/\sqrt{3}), &\qquad&
\mubf_2 = \half (1, 1, 1, 1, 1, -5/\sqrt{3}), \cr
\mubf_3 = (0, 0, 1, 1, 1, -\sqrt{3}),  &\qquad&
\mubf_4 = (0, 0, 0, 1, 1, -2/\sqrt{3}),  \cr
\mubf_5 = (0, 0, 0, 0, 1, -1/\sqrt{3}),  &\qquad&
\mubf_6 = \half (-1, 1, 1, 1, 1, -\sqrt{3}),
\eea
and $\mubf_i = \tilde\mubf_i $.

\paragraph{\color{blue}{Cartan matrix}}
Using the $\alpbf^i$ we find
\beq
a(E_6) = \bpm{
2 & -1& 0 & 0 & 0 & 0 \cr
-1& 2 & -1& 0 & 0 & 0 \cr
0 & -1& 2 & -1& 0 & -1 \cr
0 & 0 & -1& 2 & -1& 0 \cr
0 & 0 & 0 & -1& 2 & 0 \cr
0 & 0 & -1& 0 & 0 & 2} \epm,
\qquad 
a(E_6)^{-1} = \inv3\bpm{
4& 5&   6&  4& 2& 3 \cr
5& 10& 12&  8& 4& 6 \cr
6& 12& 18& 12& 6& 9 \cr
4&  8& 12& 10& 5& 6 \cr
2&  4&  6&  5& 4& 3 \cr
3&  6&  9&  6& 3& 6 } \epm,
\eeq
so that $ N = 3 $.

\paragraph{\color{blue}{$\su2$ content}}
From the Cartan matrix it follows that choosing
$ \alpbf $ to be any of the simple roots implies that all other
roots are all isodoublets or isosinglets.

\paragraph{\color{blue}{$\ui$ content}}

\bit

\item $\alpbf = \alpbf^1 $, then only $ \alpbf^2 $ is an isodoublet so that
$\hyp = h \mubf_2 $ and we find that the only non-vanishing $\hyp'_q $ are
\beq
\hyp'_{h,1/2,1/2} = \hyp'_{2h,0,0} = 3\hyp \ ,
\eeq
so there are no additional $Z'$, and (\ref{eq:sz.and.h}) requires
$ h =1/2 $; then
\beq
\alpbf = \alpbf^1 \,,\quad
\hyp = \half \mubf_2 \,,\quad
\sw^2 = \frac38 \ ,
\eeq
reminiscent of the original $\su5$ GUT.

\item $ \alpbf = \alpbf^2 $, then $ \alpbf^{1,3} $ are
isodoublets so that $ \hyp = h ( \mubf_1 + \mubf_3 ) $
and the only non-vanishing $\hyp'_q $ are
\bea
\hyp'_{1,1/2,1/2} &=& \half(0,0,3,3,3,\sqrt{3}),\cr
\hyp'_{2,0,0} = 2 \hyp'_{3,1/2,1/2}&=& (0,0,1,1,1,-3 \sqrt{3}), \cr
\hyp'_{4,0,0}&=& (0,0,1,1,1,-\sqrt{3}),
\eea
that are linear combinations of $\hyp$ and $\ee^6$,
so there will be at least one light $Z' $.

\item $ \alpbf = \alpbf^3 $, then $ \alpbf^{2,4,6} $ are
isodoublets so that $ \hyp = h ( \mubf_2 + \mubf_4 + \mubf_6 ) $
and the only non-vanishing $\hyp'_q $ are
\bea
\hyp'_{h,1/2,1/2} = \hyp'_{5h,1/2,1/2} &=& \half(1,0,0,1,1,-\sqrt{3}), \cr
\hyp'_{2h,0,0} = \hyp'_{4h,0,0}&=&\half(1,1,1,3,3,-3\sqrt{3}), \cr
\hyp'_{3h,1/2,1/2} = 2\hyp'_{6h,0,0} &=& (-1,1,1,1,1,-\sqrt{3}),
\eea
that are linear combinations of $\hyp$ and
$(1,0,0,1,1,-\sqrt{3})$, so there will be at least one light $Z' $.

\item $ \alpbf = \alpbf^4 $, then $ \alpbf^{3,5} $ are
isodoublets so that $ \hyp = h ( \mubf_3 + \mubf_5 ) $
and the only non-vanishing $\hyp'_q $ are
\bea
\hyp'_{h,1/2,1/2} &=& \half(0,0,3,3,0,-2\sqrt{3}), \cr
2\hyp'_{3h,1/2,1/2} = \hyp'_{2h,0,0} &=& (0,0,1,1,4,-2\sqrt{3}), \cr
\hyp'_{4h,0,0} &=& (0,0,1,1,1,-\sqrt{3}),
\eea
that are linear combinations of $\hyp$ and $(0,0,1,1,1,-\sqrt{3})$,
so there will be at least one light $Z' $.

\item $ \alpbf = \alpbf^5 $, then only $ \alpbf^4 $ is an
isodoublet so that $ \hyp = h \mubf_4 $
and the only non-vanishing $\hyp'_q $ are
\beq
\hyp'_{h,1/2,1/2} = \hyp'_{2h,0,0} =\mubf_4 \ ,
\eeq
so there are no light $Z' $ and (\ref{eq:sz.and.h}) requires
$ h=1/2 $; then
\beq
\alpbf = \alpbf^5 \,,\quad
\hyp = \half \mubf_4 \,,\quad
\sw^2 = \frac38 \ ,
\eeq
again reminiscent of the $\su5$ GUT.

\item $ \alpbf = \alpbf^6 $, then only $ \alpbf^3 $ is an
isodoublet so that $ \hyp = h \mubf_3 $
and the only non-vanishing $\hyp'_q $ are
\beq
\frac23 \hyp'_{h,1/2,1/2} = \inv3 \hyp'_{2h,0,0} =
2\hyp'_{3h,1/2,1/2} = \mubf_3 \ ,
\eeq
so there are no light $Z' $ and (\ref{eq:sz.and.h})
requires $h=1/2$ or $h=1/6 $ so that there
are two possibilities:
\beq
\alpbf = \alpbf^6 \,,\quad
\hyp = \half\mubf_3 \,,\quad
\sw^2 = \frac14\,; \quad {\rm or} \quad
\alpbf = \alpbf^6 \,,\quad
\hyp = \inv6 \mubf_3 \,,\quad
\sw^2 = \frac34\,.
\eeq

\eit

The cases $ \alpbf=\alpbf^1$ and $ \alpbf = \alpbf^5 $
are mapped into each other by the one non-trivial outer automorphism of $E_6$.\
(and similarly for $ \alpbf^{2,4}$).

\paragraph{\color{blue}{Matter content}}
Using
\beq
\begin{array}{|c|cccccc|}
\hline
\hyp & \hyp\cdot\mubf_1  & \hyp\cdot\mubf_2 & \hyp\cdot\mubf_3
 & \hyp\cdot\mubf_4 & \hyp\cdot\mubf_5 & \hyp\cdot\mubf_6 \cr\hline
\mubf_2/2 & 5/6 & 5/3 & 2 & 4/3 & 2/3 & 1 \cr\hline
\mubf_4/2 & 2/3 & 4/3 & 2 & 5/3 & 5/6 & 1 \cr\hline
\mubf_3/2 & 1 & 2 & 3 & 2 & 1 & 3/2 \cr\hline
\end{array} 
\eeq
together with $ \hyp\cdot\alpbf^i =0,1/2 $,
it follows that for the case  $ \alpbf = \alpbf^6$
all states will have half-integer hypercharges, and cannot
accommodate quarks.  Models for which $ \alpbf = \alpbf^{1,5} $
however, can  include quarks; for example, the {\bf27}
multiplet decomposes into
\beq
\begin{array}{|c|c|}
\hline
(h,s_z) & {\rm multiplicity} \cr \hline
(\pm1/3,0) & 10 \cr \hline
(\mp2/3,0) & 5 \cr \hline
(\mp1/6,1/2) & 5 \cr \hline
(\mp1/6,-1/2) & 5 \cr \hline
(\pm5/6,1/2) & 1 \cr \hline
(\pm5/6,-1/2) & 1 \cr \hline
\end{array}
\eeq
where the upper (lower) sign corresponds to $ \alpbf^{1 (5)}$.

\subsection{$E_7$ and $E_8$}

We followed a similar approach for the remaining exceptional groups
and found several models that can accommodate quarks and do not 
necessarily contain additional light vector bosons. Due to the
relatively large number of generators and possible choices of
$ \alpbf $ and $\hyp $ we will limit ourselves to listing
the \sm\ group embeddings for the viable models; the notation we use 
follows Gilmore's  book~\cite{Gilmore:1974}.
\bea
E_7:&& \cr
&&\hbox{case~1} ~~ \left\{
\begin{array}{l}
\alpbf = \half(1,1,1,1,1,1,\sqrt{2}) \cr
\hyp = \inv{12} (1,1,1,1,1,1,-3\sqrt{2}) \cr
\sw^2 = 3/4\cr
\end{array} \right. \cr
&&\cr
&&\hbox{case~2} ~~ \left\{
\begin{array}{l}
\alpbf=(-1,1,0,0,0,0,0) \cr
\hyp=\inv6(0,0,1,1,1,1,-2 \sqrt{2}) \cr
\sw^2=3/5 
\end{array} \right.  \cr
&&\cr
E_8:&& \cr
&&\hbox{case~1} ~~ \left\{
\begin{array}{l}
\alpbf = -\half(-1,1,1,1,1,1,1,-1)\cr
\hyp = \inv{12}(-1,1,1,1,1,1,7)\cr
\sw^2=9/16
\end{array} \right.  \cr
&&\cr
&&\hbox{case~2} ~~ \left\{
\begin{array}{l}
\alpbf = (1,1,0,0,0,0,0,0)\cr
\hyp = \inv6(0,0,1,1,1,1,1,5)\cr
\sw^2=3/8
\end{array}  \right. 
\eea

As for the other groups the values of the weak-mixing angle are much larger 
than the low-energy measurements. The smallest one is the same as the one found 
in some cases in the much simpler $E_6$ group; we do not consider this case 
further.

\end{document}